\documentclass[twoside,twocolumn,english,prl]{revtex4-1}

\usepackage[T1]{fontenc}
\usepackage[latin9]{inputenc}
\usepackage{amsmath}
\usepackage{amssymb}
\usepackage{graphicx}

\makeatletter
\@ifundefined{date}{}{\date{}}
\makeatother

\usepackage{babel}
\begin{document}

\title{Competing effects of Hund's splitting and symmetry-breaking perturbations
on electronic order in Pb$_{1-x}$Sn$_{x}$Te}

\author{Sarbajaya Kundu and Vikram Tripathi}

\affiliation{Department of Theoretical Physics, Tata Institute of Fundamental
Research, Homi Bhabha Road, Navy Nagar, Colaba, Mumbai-400005, }

\date{\today}
\begin{abstract}
We study the effect of a uniform external magnetization on $p$-wave
superconductivity on the (001) surface of the crystalline topological
insulator(TCI) Pb$_{1-x}$Sn$_{x}$Te. It was shown by us in an earlier
work that a chiral $p$-wave finite-momentum pairing (FFLO) state
can be stabilized in this system in the presence of weak repulsive
interparticle interactions. In particular, the superconducting instability
is very sensitive to the Hund's interaction in the multiorbital TCI,
and no instabilities are found to be possible for the ``wrong''
sign of the Hund's splitting. Here we show that for a finite Hund's
splitting of interactions, a significant value of the external magnetization
is needed to degrade the surface superconductivity, while in the absence
of the Hund's interaction, an arbitrarily small external magnetization
can destroy the superconductivity. This implies that multiorbital
effects in this system play an important role in stabilizing electronic
order on the surface. 
\end{abstract}
\maketitle
The topological crystalline insulator (TCI) \cite{fu2011topological}
phase is a new state of matter where the topological character of
the electronic bands is protected by crystalline symmetries. The IV-VI
semiconductors SnTe and related semiconducting alloys Pb$_{1-x}$Sn$_{x}$(Te,Se)
were recently predicted to belong to the TCI class \cite{dziawa2012topological,hsieh2012topological,tanaka2012experimental,xu2012observation}.
These have an even number of Dirac cones on high-symmetry crystal
surfaces such as $\{001\}$, $\{110\}$ and $\{111\}$, topologically
protected by the reflection symmetry with respect to the $\{110\}$
mirror planes. Here the nontrivial topology is mathematically characterized
by a mirror Chern number \cite{hsieh2012topological}, and topologically
protected surface states with novel electronic dispersions are present
on the different surfaces invariant under reflection symmetry. In
particular, it has been shown in \cite{liu2013two} that the (001)
surface of Pb$_{1-x}$Sn$_{x}$Te comprises two disconnected Dirac
pockets touching each other at two saddle points, giving rise to Type-II
Van-Hove singularities \cite{PhysRevB.92.035132} in the density of
states. This enhances the possibility of competing Fermi-surface instabilities
on the TCI surface, brought about by weak repulsive interparticle
interactions \cite{dzyaloshinskii1987maximal,dzyaloshinskii1996extended,baranov1992superconductivity,bernevig2006quantum,gonzalez1996renormalization,honerkamp2001magnetic,mcchesney2010extended}.

In order to study the competition between different electronic orders,
a weak-coupling renormalization group analysis, which treats all the
competing orders on an equal footing, is desirable. Such a parquet
approximation for systems with multiple Fermi pockets has proved useful
in the past for studies on unconventional superconductivity \cite{norman2011challenge,mineev1999introduction,sigrist1991phenomenological}
in multiple other systems such as cuprates \cite{furukawa1998truncation},
graphene \cite{nandkishore2012chiral} and semimetal thin films \cite{PhysRevB.93.155108}.
In the presence of Fermi surface nesting, charge and spin density
wave orders tend to compete with superconductivity \cite{le2009superconductivity}.
In a multiorbital system like Pb$_{1-x}$Sn$_{x}$Te, one also has
to take into account the effect of Hund's splitting of repulsive electron
interactions. We have shown in an earlier work \cite{kundu2017role}
that the electrons interacting via repulsive interactions on the TCI
surface are unstable against a chiral $p$-wave superconducting order,
where the Van-Hove singularities serve to enhance the effective transition
temperature and the (approximate) nesting allows the particle-hole
instabilities to compete with superconductivity on an equal footing.
Interestingly, here the $p$-wave symmetry arises not from intrinsic
Fermi-surface deformations but from the nontrivial Berry phases associated
with the topological surface states. Moreover, the very existence
of the superconducting state is sensitive to the Hund's interaction
in this system and no instabilities are found to occur for a negative
Hund's splitting. Due to the presence of the low-lying Van-Hove singularities
on the TCI surface, such a state promises to be experimentally accessible. 

A relevant question which ought to be addressed in this context is:
how robust is such a superconducting order against a time-reversal
symmetry breaking perturbation, such as proximity coupling to an external
magnetization. The effects of various symmetry-breaking perturbations,
including a perpendicular magnetic field or moments, on the surface
states of the TCI (as well as other topological insulators) have been
studied both theoretically and experimentally \cite{article,huang2016hedgehog,liu2013spin,PhysRevLett.112.046801,xu2012hedgehog};
however, here we focus on an aspect of the system that has been previously
overlooked. We show that the robustness of the surface superconducting
order against an external magnetization is enhanced by the presence
of a finite Hund's interaction. Specifically, the critical value of
spin-splitting (induced by the magnetization), beyond which $p$-wave
superconductivity is no longer possible, scales directly with the
size of the Hund's splitting with respect to the repulsive electron
interaction strength. This implies that multiorbital effects in this
system play an important role in stabilizing electronic order on the
surface. 

The rest of the paper is organized as follows. In Sec. 1,
we introduce the k.p Hamiltonian for the (001) surface and describe
some of the features of eigenstates and the spectrum in the presence
of a spin-splitting term. Sec. 2 presents the low-energy
theory for weakly repulsive electronic interactions, including a finite
Hund's splitting. In Sec. 3, the parquet renormalization
group equations for the couplings are provided. In Sec. 4,
we solve the renormalization group equations and obtain the singular
behavior of different susceptibilities near the critical point. Sec.
5 is devoted to situations where the parquet analysis is
no longer valid (if the parquet fixed point corresponds to an energy
scale below the Fermi surface) for which we perform the usual ladder
renormalization group analysis and obtain the instabilities in the
system. In Sec. 6, we determine the behavior of the critical
values of spin-splitting below which $p$-wave superconductivity
is stable in the presence of a finite Hund's interaction,
as a function of the Hund's splitting introduced. Sec. 7
contains a summary of our results and a discussion.

\section*{1. \label{sec1}The surface hamiltonian and two-dimensional van-hove
singularities}

\begin{figure*}
(a)\includegraphics[width=1.0\columnwidth]{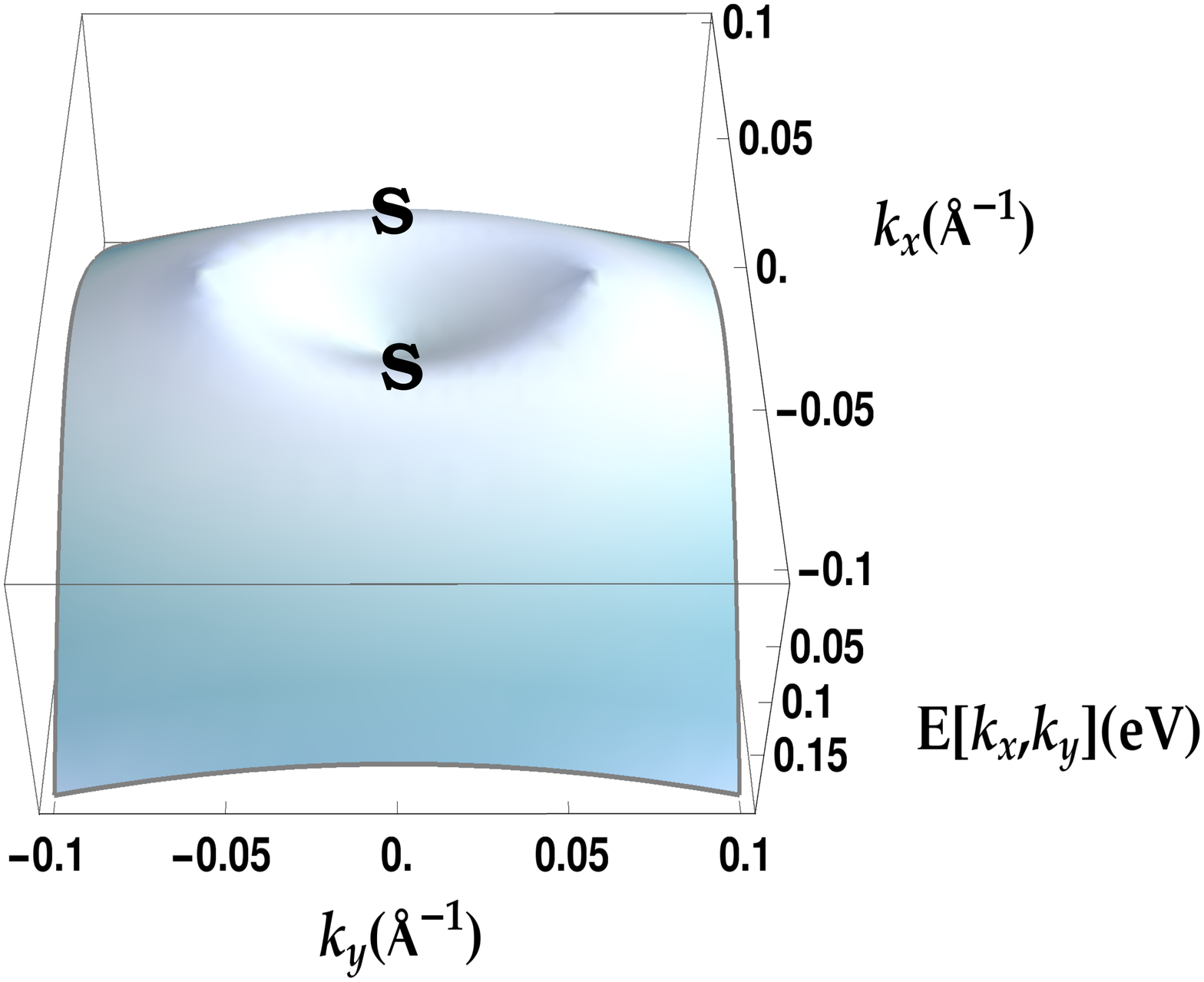}(b)\includegraphics[width=1.0\columnwidth]{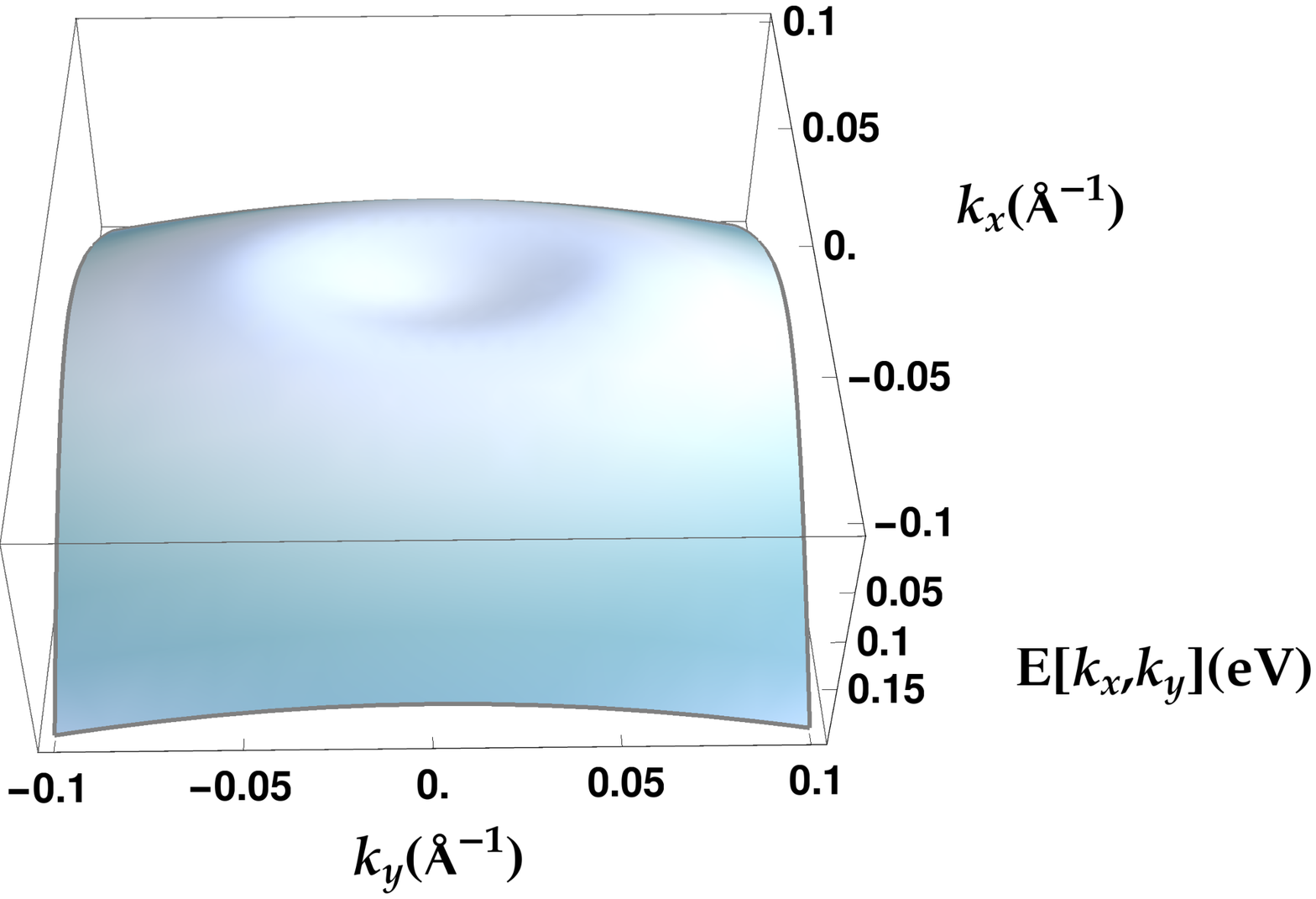}

(c)\includegraphics[width=1.0\columnwidth]{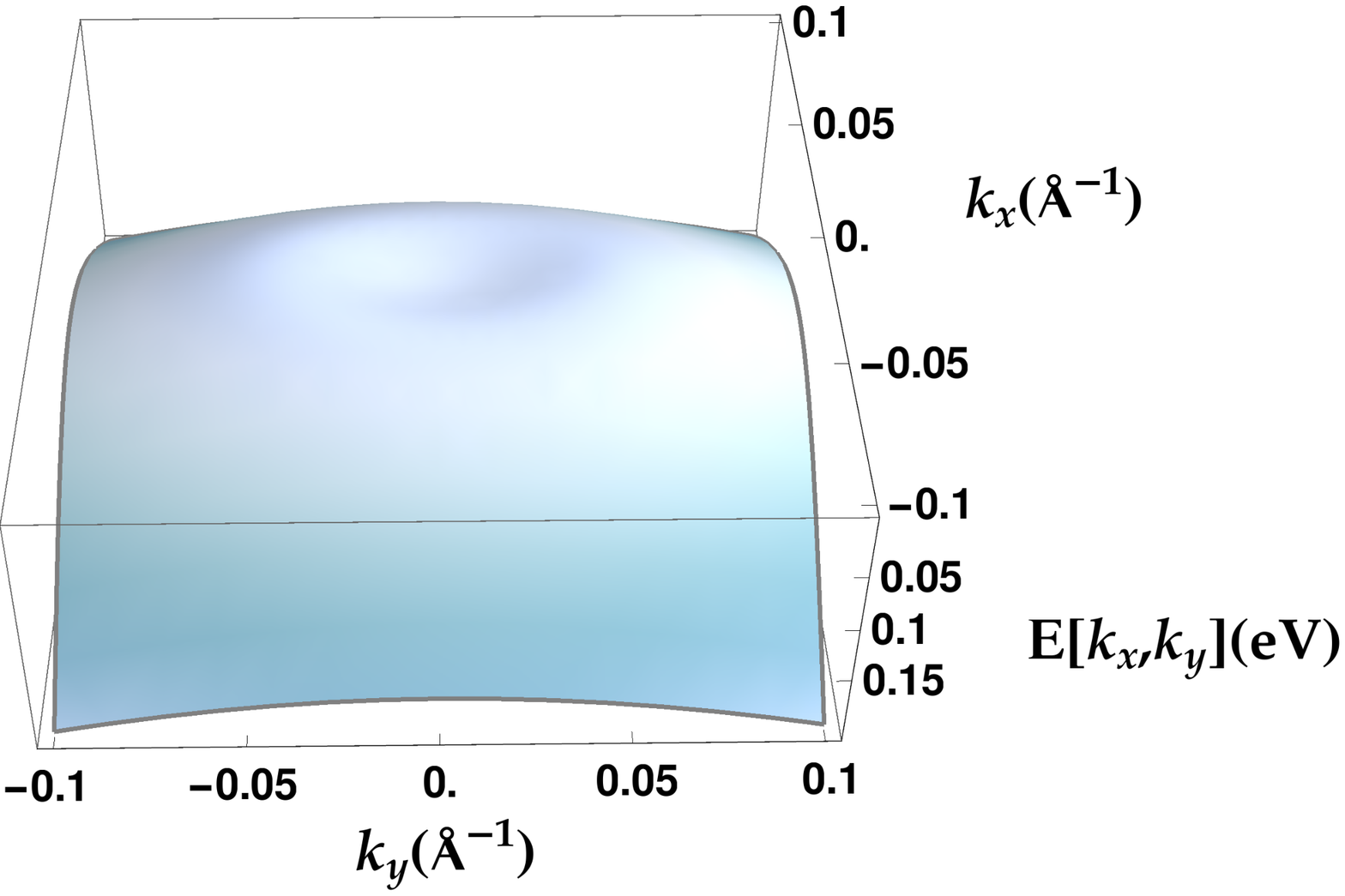}(d)\includegraphics[width=1.0\columnwidth]{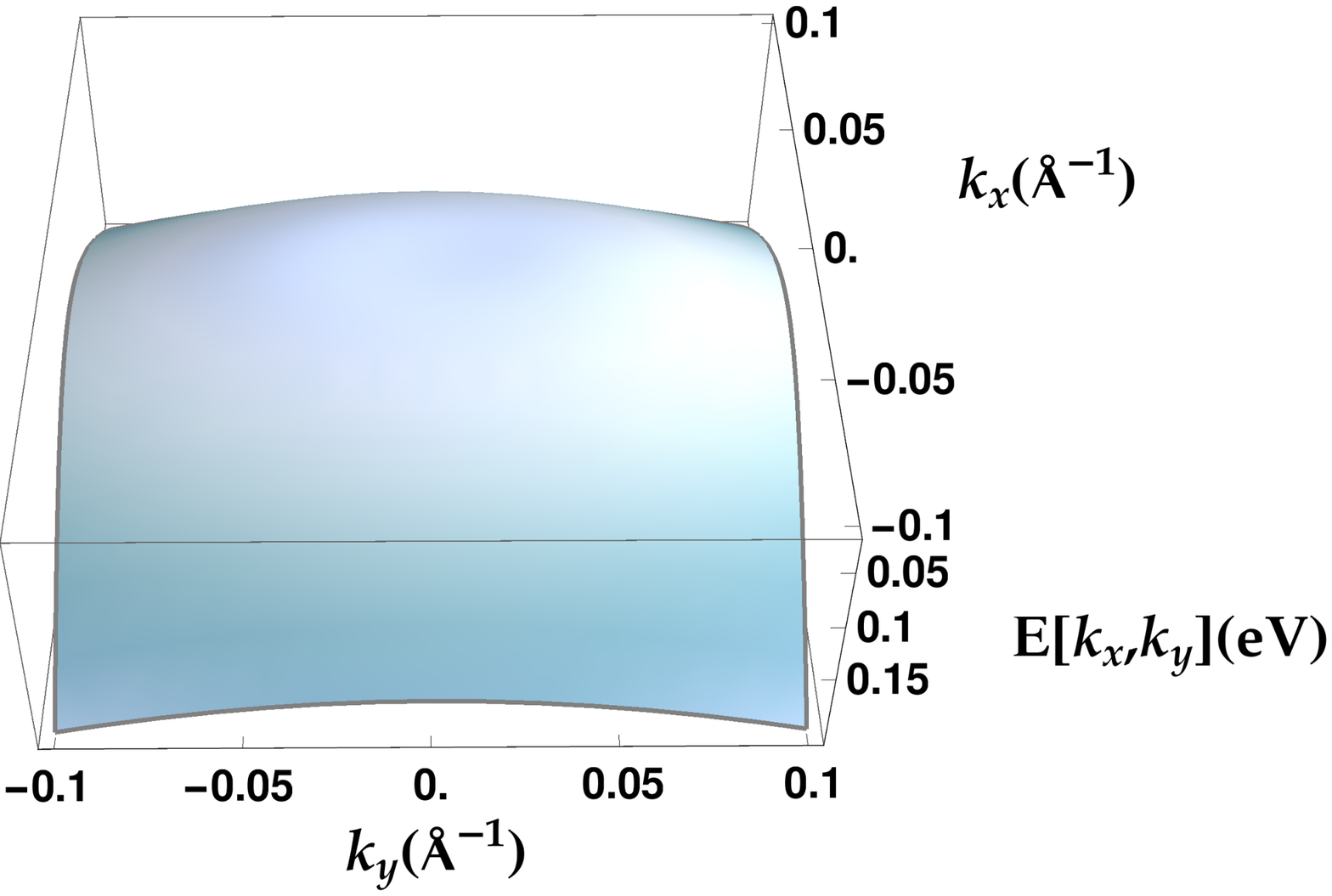}

\caption{\label{fig:The-two-dimensional-Van-Hove}The two-dimensional Van-Hove
singularities on the TCI surface (indicated by 'S') for (a) $M=0.0$,
(b) $M=0.04$, (c) $M=0.05$, and (d) $M=0.07$ (in eV) where $v_{x}=2.4$
eV $\mathring{A}^{-1}$, $v_{y}=1.3$ eV $\mathring{A}^{-1}$, $m=0.07$
eV and $\delta=0.026$ eV (values taken from \cite{liu2013two}).
We find that beyond $M\approx0.05$ eV, there are no Van-Hove singularities
in the surface electronic spectrum. }
\end{figure*}

\begin{figure*}
(a)\includegraphics[width=1.0\columnwidth]{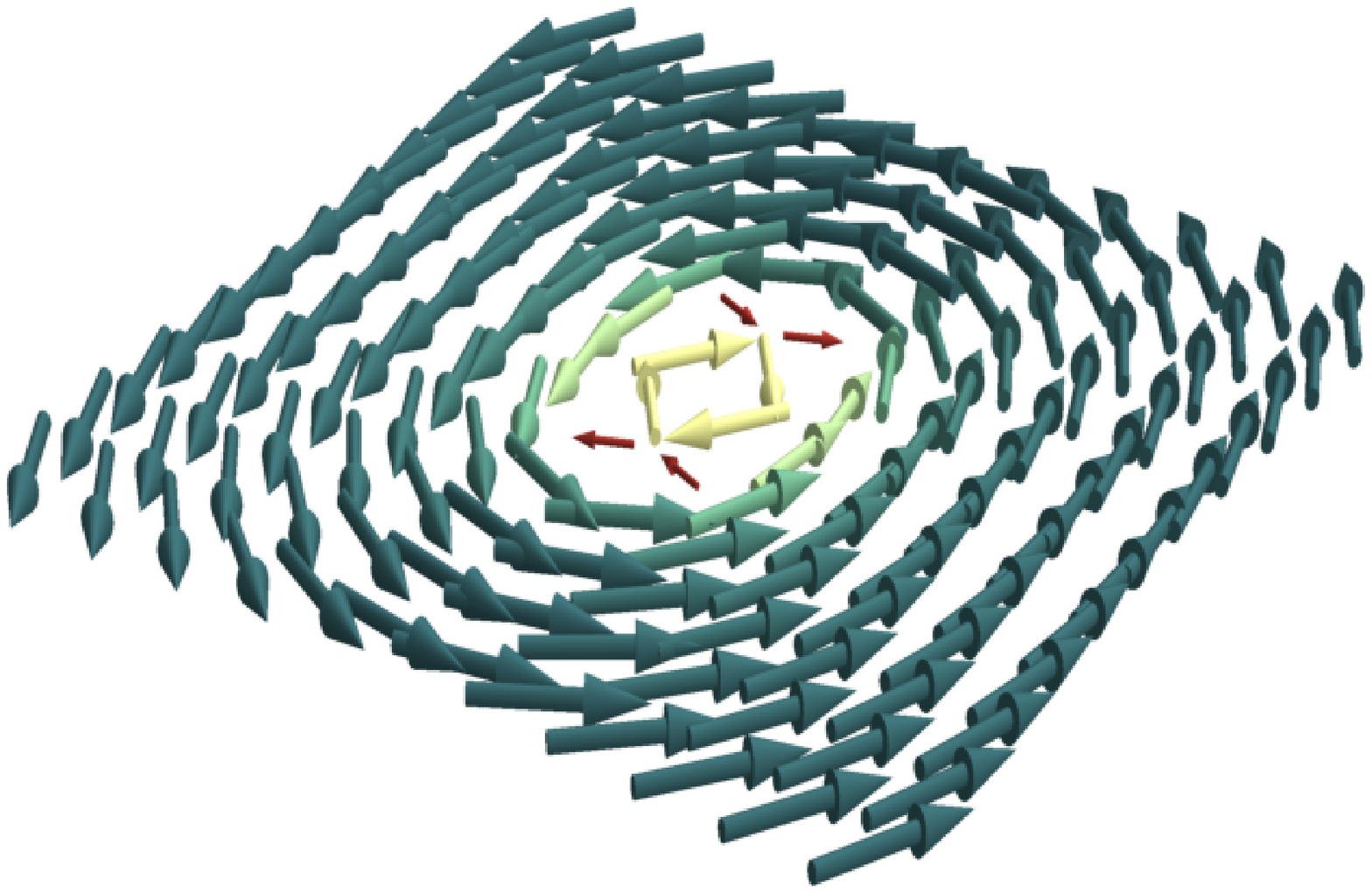}(b)\includegraphics[width=1.0\columnwidth]{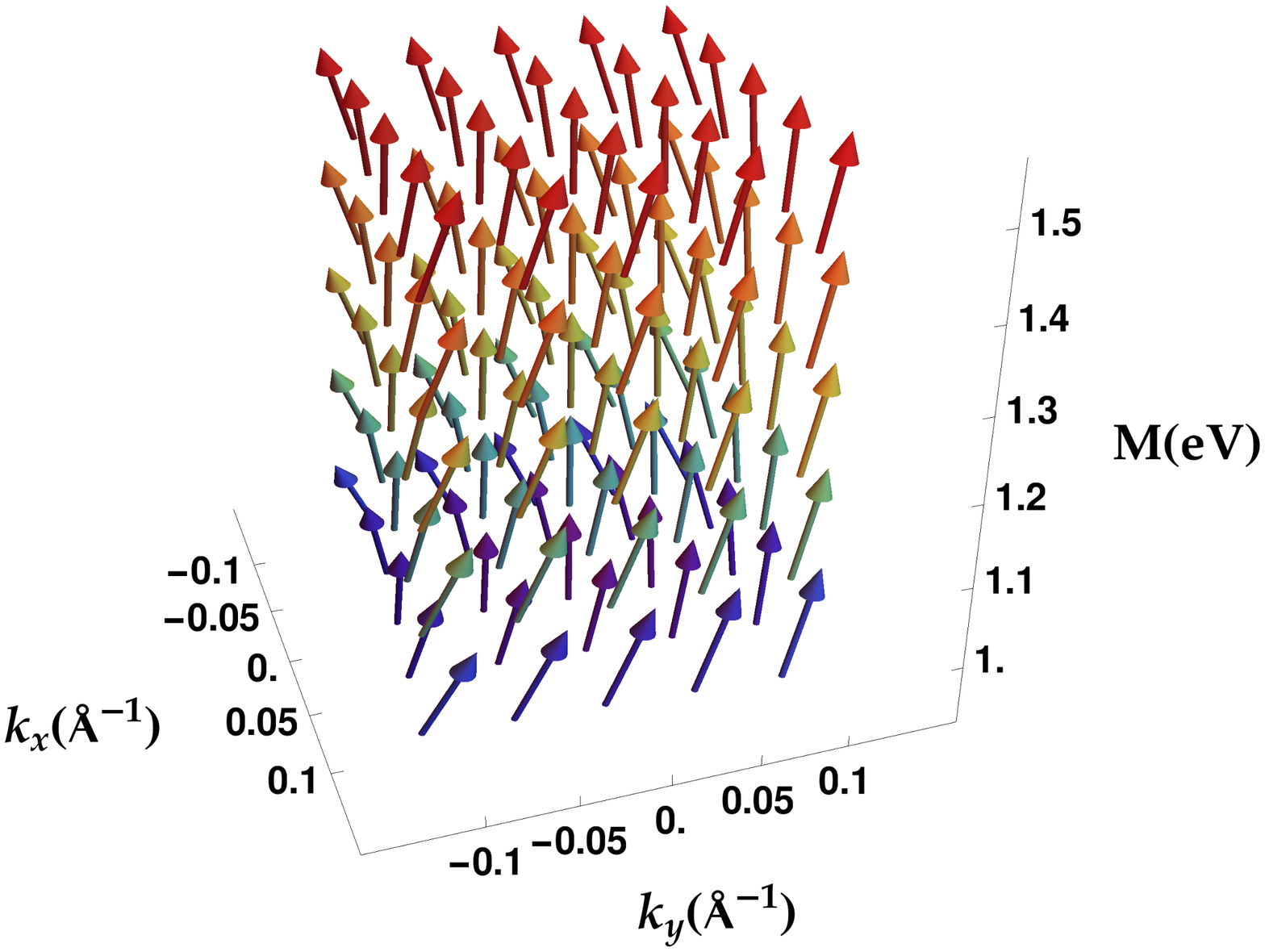}

\caption{\label{fig:chiralspin}The spin texture $<\protect\overrightarrow{s}(k_{x},k_{y})>=(<s_{x}>,<s_{y}>,<s_{z}>)$
on the TCI surface (where $k_{x}$ and $k_{y}$ are measured in $\mathring{A}^{-1}$)
for the band with a positive energy lying closest to the Fermi level,
with (a) $M=0.0$ eV, and (b) $M$ in the range $1$ to $1.5$ eV.
When $M$ is small, the TCI surface continues to have a chiral spin
texture, shown in (a), and for larger values of $M$, the spins gradually
develop an out-of-plane polarization.}
\end{figure*}

The fundamental band gaps of IV-VI semiconductors are located at the
four equivalent $L$ points in the FCC Brillouin zone. According to
\cite{liu2013two}, the TCI surface states can be classified into
two types: \emph{Type-I}, for which all four $L$-points are projected
to the different time-reversal invariant momenta(TRIM) in the surface
Brillouin zone, and \emph{Type-II}, for which different $L$-points
are projected to the same surface momentum. The (001) surface falls
into the latter class of surfaces. Here, the plane $\Gamma L_{1}L_{2}$
in the bulk Brillouin zone projects onto the line $\overline{\Gamma}\overline{X_{1}}$
on the surface, such that $L_{1}$ and $L_{2}$ both project onto
the $\overline{X_{1}}$ point. Similarly, $L_{3}$ and $L_{4}$ project
onto the symmetry-related point $\overline{X_{2}}$. This leads to
two coexisting massless Dirac fermions at $\overline{X_{1}}$ arising
from the $L_{1}$ and the $L_{2}$ valley, respectively, and likewise
at $\overline{X_{2}}$. The k.p Hamiltonian close to the point $\overline{X_{1}}$
on the (001) surface is derived on the basis of a symmetry analysis
in \cite{liu2013two}, and is given by 
\begin{equation}
H_{\overline{X_{1}}}(k)=(v_{x}k_{x}s_{y}-v_{y}k_{y}s_{x})+m\tau_{x}+\delta s_{x}\tau_{y},\label{eq:1}
\end{equation}
where $k$ is measured with respect to $\overline{X_{1}}$, $\tau$
operates in valley ($L$) space and $\overrightarrow{s}$ is a set
of Pauli matrices associated with the two spin components associated
with each valley, and the terms $m$ and $\delta$ are added to describe
intervalley scattering. The band dispersion and constant energy contours
for the above surface Hamiltonian undergo a Lifshitz transition with
increasing energy away from the Dirac point, and when the Fermi surface
is at $\sim26$ meV \cite{liu2013two}, two saddle points $\overline{S_{1}}$
and $\overline{S_{2}}$ at momenta $(\pm\frac{m}{v_{x}},0)$ lead
to a Van-Hove singularity in the density of states. A similar situation
arises at the point $\overline{X_{2}}$. This is illustrated in Fig.
\ref{fig:The-two-dimensional-Van-Hove}(a). Here we further introduce
a spin-splitting term $Ms_{z}$ into the surface Hamiltonian in Eq.
\ref{eq:1}, which breaks the degeneracy between the two spin components.
The authors of ref. \cite{article} have also incorporated additional
terms (involving the valley degrees of freedom) to describe the Zeeman
coupling of the TCI surface states to a perpendicular magnetic field
or magnetic moment, deduced on the basis of a symmetry analysis of
the surface Hamiltonian. We have repeated our calculations for their
model and find that the results change only quantitatively, and hence,
we shall consider the simpler case with a Zeeman-only perturbation.
We find that the Van-Hove singularities in the surface bandstructure
survive up to $M\approx0.05$ eV (with parameters suitable for SnTe
being taken from \cite{liu2013two}), when the Fermi level is at $\approx0.026$
eV, the position of the Van-Hoves in the absence of $M$. The evolution
of the two-dimensional Van-Hove singularities with an increase in
the Zeeman splitting $M$ is shown in the Fig. \ref{fig:The-two-dimensional-Van-Hove}.
The TCI surface is found to have a chiral spin texture in the absence
of an external magnetization, and as $M$ (>0) increases, it acquires
an out-of-plane spin polarization. This is graphically depicted in
Fig. \ref{fig:chiralspin}. The topological defects are visible at
the $\overline{X}$ point. 

\section*{2. \label{sec2}Electron interactions in the valley-spin basis and
form factors}

In addition to the noninteracting Hamiltonian with an external magnetization,
described in the previous section, we also consider interactions between
electrons on different $L$-valleys for different spin combinations.
We then project the interactions in the valley-spin basis to the positive
energy band lying closest to the Van-Hove singularities (corresponding
to each of the $\overline{X}$ points) where the Fermi level is fixed,
and this gives rise to additional form factors in the different interaction
terms. We find that the form factors $u_{a\uparrow}$ corresponding
to the spin $\uparrow$ component (for each valley $a$) have an additional
phase factor of $\exp[i\theta_{k}]$ where $\theta_{k}$ is measured
with respect to the $\overline{X}$ point, and the argument of the
form factors $u_{a\downarrow}$ for the spin $\downarrow$ components
do not change upon advancing by $2\pi$ around the $\overline{X}$
point. Upon substituting the appropriate form factors into the electron
interaction model in the valley-spin basis, the low-energy theory
for the effective interaction model in the band picture is given by
(please refer to \cite{kundu2017role} for details) 
\begin{align}
L & =\sum_{i,\sigma,\sigma^{\prime}}\left[\psi_{i}^{\dagger}(\partial_{\tau}-\epsilon_{k}+\mu)\psi_{i}-\frac{1}{2}h_{4}^{\sigma\sigma^{\prime}}\psi_{i}^{\dagger}\psi_{i}^{\dagger}\psi_{i}\psi_{i}\right.\nonumber \\
 & \qquad-\sum_{i\neq j}\frac{1}{2}(h_{1}^{\sigma\sigma^{\prime}}\psi_{i}^{\dagger}\psi_{j}^{\dagger}\psi_{i}\psi_{j}+h_{2}^{\sigma\sigma^{\prime}}\psi_{i}^{\dagger}\psi_{j}^{\dagger}\psi_{j}\psi_{i}\nonumber \\
 & \qquad\left.+h_{3}^{\sigma\sigma^{\prime}}\psi_{i}^{\dagger}\psi_{i}^{\dagger}\psi_{j}\psi_{j})\right],\label{eq:5}
\end{align}
where the summation $i$ is over the two bands (corresponding to the
two points $\overline{X_{1}}$ and $\overline{X_{2}}$) being considered.
Here, $\epsilon_{k}$ is obtained by diagonalizing Eq. \ref{eq:1},
along with the spin-splitting term $Ms_{z}$. The chemical potential
value $\mu$=0 corresponds to the system being doped to the Van Hove
singularities. Here $h_{4}$ refers to scattering processes between
different valleys within a band $i$. The couplings $h_{1}$, $h_{2}$
and $h_{3}$ represent exchange effects, Coulomb interaction and pair
hopping, respectively, between electrons in different bands. Due to
the distinctive phase dependences of the form factors, the interactions
which correspond to spin-antiparallel configurations have an additional
phase dependence of $\exp[i(\theta_{k}-\theta_{k^{\prime}})]$ and
transform as $\ell=1$ functions in 2D, while those corresponding
to spin-parallel configurations transform as $\ell=0$ functions.
This implies that Coulomb interactions between the surface electrons
generally depend on their relative spin configuration, and we therefore
distinguish between interactions between electrons with parallel and
anti-parallel spin configurations in our analysis. In the absence
of a spin-splitting term in the Hamiltonian, the momentum-dependence
of the interactions can be incorporated entirely into the aforementioned
phase factors. In our earlier analysis in \cite{kundu2017role}, we
have neglected the momentum-dependence of the absolute values of the
form factors $u_{\uparrow,\downarrow}$(for each of the valleys).
However, in the presence of a spin-splitting term $M$, the degeneracy
between spins $\uparrow$ and $\downarrow$ is broken, and the complex
form factors $u_{\uparrow}$ and $u_{\downarrow}$ differ both in
amplitude and phase.

\section*{3. \label{sec3}Renormalization group equations in the presence of
spin-splitting}

\begin{figure}
\includegraphics[width=1\columnwidth]{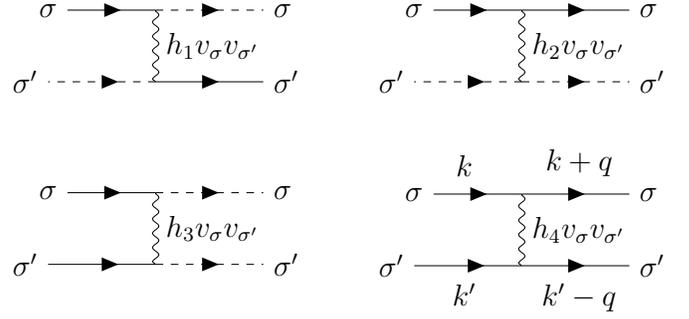}

\caption{\label{fig:The-RG-couplings}Couplings $h_{i}$ defined on a patch($h_{4}$)
and between the two patches($h_{1}$,$h_{2}$,$h_{3}$). There is
a momentum-dependence associated with all of these couplings. Each
of the possible scattering processes within a patch are denoted by
$h_{4}$ in our analysis. The spin labels $\sigma,\sigma^{\prime}$
correspond to specific spin components of the spinor wavefunctions
associated with the different bands under consideration. The nature
of the couplings considered here are very similar to those defined
in \cite{kundu2017role}, except now each of them also has an explicit
factor $v_{\sigma}v_{\sigma^{\prime}}$depending on the particular
spin combination being considered.}
\end{figure}

We perform our RG analysis with Fermi patches located at the two points
$\overline{X_{1}}$ and $\overline{X_{2}}$ on the (001) surface,
near the Van Hove singularities. As mentioned earlier, the phases
arising from the form factors distinguish between electron interactions
with parallel and antiparallel spin configurations, and so each of
the couplings $h_{i}$ (for the scattering channel $i=1-4$) in our
RG analysis now have two components $h_{i}^{\sigma\sigma}$ and $h_{i}^{\sigma\overline{\sigma}}$,
which we shall denote as $h_{i}^{0}$ (for $\ell=0$) and $h_{i}^{1}$
(for $\ell=1$) respectively. Moreover, the two spin components $\uparrow$
and $\downarrow$ are inequivalent in the presence of the Zeeman splitting
term, and this gives rise to additional components for the couplings.
To simplify our analysis, we have integrated out the momentum-dependence
of the absolute values of the form factors $|u_{\uparrow}(\overrightarrow{k},M)|^{2}$
and $|u_{\downarrow}(\overrightarrow{k},M)|^{2}$ for the two spin
components, over a suitable range of two-dimensional momenta $(k_{x},k_{y})$
around the $\overline{X}$ points (chosen to be $0.1$ $\mathring{A}^{-1}$
for our calculations, based on the scale associated with the dispersion
in momentum space as shown in Fig. \ref{fig:The-two-dimensional-Van-Hove}),
and normalized the results with respect to $|u_{\uparrow}(\overrightarrow{k},0)|^{2}$
and $|u_{\downarrow}(\overrightarrow{k},0)|^{2}$ respectively. Henceforth,
we shall denote these $k$-integrated form factors by $v_{\uparrow}$
and $v_{\downarrow}$ for simplicity. The couplings constants $h_{i}$
associated with the RG flows either involve two factors of either
$v_{\uparrow}$ or $v_{\downarrow}$, or one factor of each. Clearly,
for $M>0$, we have $v_{\uparrow}(M)>1$ and $v_{\downarrow}(M)<1$
for the positive energy eigenstates, and the ratio $\frac{v_{\uparrow}(M)}{v_{\downarrow}(M)}$
increases with an increase in $M$. Corresponding to every scattering
channel $h_{i}$, we then have four components $h_{i}^{\uparrow\uparrow}$,
$h_{i}^{\downarrow\downarrow}$, $h_{i}^{\uparrow\downarrow}$ and
$h_{i}^{\downarrow\uparrow}$, alternately denoted by $h_{i}^{0}$,
$h_{i}^{2}$,$h_{i}^{1}$ and $h_{i}^{3}$ respectively. This gives
us a set of 16 coupling constants. The different coupling constants
for interactions within a patch as well as between patches are shown
in Fig. \ref{fig:The-RG-couplings}, taking into account the explicit
factors of $v_{\uparrow}$ and $v_{\downarrow}$. For $M\gtrsim0.05$
eV, one also has to take into account the absence of the Van-Hove
singularities in the spectrum. We have performed calculations for
higher values of magnetization as well, and found that the qualitative
behavior of the system in that regime is very similar to what we discuss
below. Therefore, we confine our attention to situations where Van-Hove
singularities are present, since that gives us high transition temperatures
even in the weak-coupling regime. 

We perform RG analysis up to one-loop level, integrating out high-energy
degrees of freedom gradually from an energy cutoff $\Lambda$, which
is the bandwidth. The susceptibilities in the different channels schematically
behave as $\chi_{0}^{pp}(\omega)\sim\ln[\Lambda/\omega]\ln[\Lambda/\text{max}(\omega,\mu)]$,
$\chi_{Q}^{ph}(\omega)\sim\ln[\Lambda/\text{max}(\omega,\mu)]\ln[\Lambda/\text{max}(\omega,\mu,t)]$
and $\chi_{0}^{ph}(\omega),\chi_{Q}^{pp}(\omega)\sim\ln[\Lambda/\text{max}(\omega,\mu)]$,
where $\omega$ denotes the energy away from the Van Hove singularities
and $t$ represents terms in the Hamiltonian that destroy the perfect
nesting. In what follows, we shall use $y\equiv\ln^{2}[\Lambda/\omega]\sim\chi_{0}^{pp}$
as the RG flow parameter, and describe the relative weight of the
other channels as $d_{1}(y)=\frac{d\chi_{Q}^{ph}}{dy}$, $d_{2}(y)=\frac{d\chi_{0}^{ph}}{dy}$
and $d_{3}(y)=-\frac{d\chi_{Q}^{pp}}{dy}$. The factor $d_{1}(y)$,
which incorporates the effects of imperfect nesting, is taken to be
a function $\frac{1}{\sqrt{1+y}}$ \cite{nandkishore2012chiral},
interpolating smoothly in between the limits $d_{1}(y=0)=1$ and $d_{1}(y\gg1)=\frac{1}{\sqrt{y}}$.
We also assume that $d_{2},d_{3}\ll d_{1}$, and neglect the terms
in the RG equations with these coefficients. The RG equations are
obtained by evaluating second-order diagrams and collecting the respective
combinatoric prefactors, for each of the interactions $h_{1}$,$h_{2}$,$h_{3}$
and $h_{4}$. The diagrams for the coupling $h_{2}$ are pictorially
illustrated in \cite{kundu2017role}. The final set of RG equations
obtained by taking into account the multiplicative factors $v_{\sigma}$
and $v_{\overline{\sigma}}$, are given by
\begin{eqnarray}
\frac{dh_{1}^{\sigma\sigma}}{dy} & = & \frac{2}{\sqrt{1+y}}((h_{3}^{\sigma\sigma})^{2}v_{\sigma}^{2}\nonumber \\
 &  & -(h_{1}^{\sigma\sigma})^{2}v_{\sigma}^{2}-(h_{3}^{\sigma\overline{\sigma}})(h_{3}^{\overline{\sigma}\sigma})v_{\overline{\sigma}}^{2}\nonumber \\
 &  & -(h_{1}^{\sigma\overline{\sigma}})(h_{1}^{\overline{\sigma}\sigma})v_{\overline{\sigma}}^{2}+2(h_{1}^{\sigma\sigma})(h_{2}^{\sigma\sigma})v_{\sigma}^{2})\qquad\label{eq:6}
\end{eqnarray}
\begin{eqnarray}
\frac{dh_{1}^{\overline{\sigma}\overline{\sigma}}}{dy} & = & \frac{2}{\sqrt{1+y}}((h_{3}^{\overline{\sigma}\overline{\sigma}})^{2}v_{\overline{\sigma}}^{2}\nonumber \\
 &  & -(h_{1}^{\overline{\sigma}\overline{\sigma}})^{2}v_{\overline{\sigma}}^{2}-(h_{3}^{\overline{\sigma}\sigma})(h_{3}^{\sigma\overline{\sigma}})v_{\sigma}^{2}\nonumber \\
 &  & -(h_{1}^{\overline{\sigma}\sigma})(h_{1}^{\sigma\overline{\sigma}})v_{\sigma}^{2}+2(h_{1}^{\overline{\sigma}\overline{\sigma}})(h_{2}^{\overline{\sigma}\overline{\sigma}})v_{\overline{\sigma}}^{2})\qquad\label{eq:7}
\end{eqnarray}
\begin{eqnarray}
\frac{dh_{1}^{\sigma\overline{\sigma}}}{dy} & = & \frac{2h_{1}^{\sigma\overline{\sigma}}}{\sqrt{1+y}}(-(h_{1}^{\overline{\sigma}\overline{\sigma}})v_{\overline{\sigma}}^{2}-(h_{1}^{\sigma\sigma})v_{\sigma}^{2}\nonumber \\
 &  & +(h_{2}^{\sigma\sigma})v_{\sigma}^{2}+(h_{2}^{\overline{\sigma}\overline{\sigma}})v_{\overline{\sigma}}^{2})\qquad\qquad\qquad\qquad\label{eq:8}
\end{eqnarray}
\begin{eqnarray}
\frac{dh_{1}^{\overline{\sigma}\sigma}}{dy} & = & \frac{2h_{1}^{\overline{\sigma}\sigma}}{\sqrt{1+y}}(-(h_{1}^{\overline{\sigma}\overline{\sigma}})v_{\overline{\sigma}}^{2}-(h_{1}^{\sigma\sigma})v_{\sigma}^{2}\nonumber \\
 &  & +(h_{2}^{\sigma\sigma})v_{\sigma}^{2}+(h_{2}^{\overline{\sigma}\overline{\sigma}})v_{\overline{\sigma}}^{2})\qquad\qquad\qquad\qquad\label{eq:9}
\end{eqnarray}
\begin{equation}
\frac{dh_{2}^{\sigma\sigma}}{dy}=\frac{2}{\sqrt{1+y}}((h_{2}^{\sigma\sigma})^{2}+(h_{3}^{\sigma\sigma})^{2})v_{\sigma}^{2}\qquad\qquad\label{eq:10}
\end{equation}
\begin{equation}
\frac{dh_{2}^{\overline{\sigma}\overline{\sigma}}}{dy}=\frac{2}{\sqrt{1+y}}((h_{2}^{\overline{\sigma}\overline{\sigma}})^{2}+(h_{3}^{\overline{\sigma}\overline{\sigma}})^{2})v_{\overline{\sigma}}^{2}\qquad\qquad\label{eq:11}
\end{equation}
\begin{equation}
\frac{dh_{2}^{\sigma\overline{\sigma}}}{dy}=\frac{2}{\sqrt{1+y}}((h_{2}^{\sigma\overline{\sigma}})^{2}+(h_{3}^{\sigma\overline{\sigma}})^{2})v_{\sigma}\,v_{\overline{\sigma}}\qquad\qquad\label{eq:12}
\end{equation}
\begin{equation}
\frac{dh_{2}^{\overline{\sigma}\sigma}}{dy}=\frac{2}{\sqrt{1+y}}((h_{2}^{\overline{\sigma}\sigma})^{2}+(h_{3}^{\overline{\sigma}\sigma})^{2})v_{\sigma}v_{\overline{\sigma}}\qquad\label{eq:13}
\end{equation}
\\
\begin{eqnarray}
\frac{dh_{3}^{\sigma\sigma}}{dy} & = & -4h_{3}^{\sigma\sigma}h_{4}^{\sigma\sigma}v_{\sigma}^{2}+\frac{2}{\sqrt{1+y}}(4h_{2}^{\sigma\sigma}h_{3}^{\sigma\sigma}v_{\sigma}^{2}\nonumber \\
 &  & -h_{1}^{\overline{\sigma}\sigma}h_{3}^{\sigma\overline{\sigma}}v_{\overline{\sigma}}^{2}-h_{1}^{\sigma\overline{\sigma}}h_{3}^{\overline{\sigma}\sigma}v_{\overline{\sigma}}^{2})\qquad\qquad\qquad\label{eq:14}
\end{eqnarray}
\begin{eqnarray}
\frac{dh_{3}^{\overline{\sigma}\overline{\sigma}}}{dy} & = & -4h_{3}^{\overline{\sigma}\overline{\sigma}}h_{4}^{\overline{\sigma}\overline{\sigma}}v_{\overline{\sigma}}^{2}+\frac{2}{\sqrt{1+y}}(4h_{2}^{\overline{\sigma}\overline{\sigma}}h_{3}^{\overline{\sigma}\overline{\sigma}}v_{\overline{\sigma}}^{2}\nonumber \\
 &  & -h_{1}^{\overline{\sigma}\sigma}h_{3}^{\sigma\overline{\sigma}}v_{\sigma}^{2}-h_{1}^{\sigma\overline{\sigma}}h_{3}^{\overline{\sigma}\sigma}v_{\sigma}^{2})\qquad\qquad\qquad\label{eq:15}
\end{eqnarray}
\begin{eqnarray}
\frac{dh_{3}^{\sigma\overline{\sigma}}}{dy} & = & -4h_{3}^{\sigma\overline{\sigma}}h_{4}^{\sigma\overline{\sigma}}v_{\sigma}v_{\overline{\sigma}}\nonumber \\
 &  & +\frac{2}{\sqrt{1+y}}(2h_{2}^{\sigma\overline{\sigma}}h_{3}^{\sigma\overline{\sigma}}v_{\sigma}v_{\overline{\sigma}}+\nonumber \\
 &  & h_{3}^{\sigma\overline{\sigma}}(h_{2}^{\sigma\sigma}v_{\sigma}^{2}+h_{2}^{\overline{\sigma}\overline{\sigma}}v_{\overline{\sigma}}^{2}-h_{1}^{\sigma\sigma}v_{\sigma}^{2}-h_{1}^{\overline{\sigma}\overline{\sigma}}v_{\overline{\sigma}}^{2}))
\end{eqnarray}
\begin{eqnarray}
\frac{dh_{3}^{\overline{\sigma}\sigma}}{dy} & = & -4h_{3}^{\overline{\sigma}\sigma}h_{4}^{\overline{\sigma}\sigma}v_{\sigma}v_{\overline{\sigma}}\nonumber \\
 &  & +\frac{2}{\sqrt{1+y}}(2h_{2}^{\overline{\sigma}\sigma}h_{3}^{\overline{\sigma}\sigma}v_{\sigma}v_{\overline{\sigma}}+\nonumber \\
 &  & h_{3}^{\overline{\sigma}\sigma}(h_{2}^{\sigma\sigma}v_{\sigma}^{2}+h_{2}^{\overline{\sigma}\overline{\sigma}}v_{\overline{\sigma}}^{2}-h_{1}^{\sigma\sigma}v_{\sigma}^{2}-h_{1}^{\overline{\sigma}\overline{\sigma}}v_{\overline{\sigma}}^{2}))
\end{eqnarray}
\begin{equation}
\frac{dh_{4}^{\sigma\sigma}}{dy}=-2((h_{3}^{\sigma\sigma})^{2}+(h_{4}^{\sigma\sigma})^{2})v_{\sigma}^{2}\qquad\qquad\qquad\label{eq:18}
\end{equation}
\begin{equation}
\frac{dh_{4}^{\overline{\sigma}\overline{\sigma}}}{dy}=-2((h_{3}^{\overline{\sigma}\overline{\sigma}})^{2}+(h_{4}^{\overline{\sigma}\overline{\sigma}})^{2})v_{\overline{\sigma}}^{2}\qquad\qquad\qquad\label{eq:19}
\end{equation}
\begin{equation}
\frac{dh_{4}^{\sigma\overline{\sigma}}}{dy}=-2((h_{3}^{\sigma\overline{\sigma}})^{2}+(h_{4}^{\sigma\overline{\sigma}})^{2})v_{\sigma}v_{\overline{\sigma}}\qquad\qquad\qquad\label{eq:20}
\end{equation}
\begin{equation}
\frac{dh_{4}^{\overline{\sigma}\sigma}}{dy}=-2((h_{3}^{\overline{\sigma}\sigma})^{2}+(h_{4}^{\overline{\sigma}\sigma})^{2})v_{\sigma}v_{\overline{\sigma}}\qquad\qquad\qquad\label{eq:21}
\end{equation}
where the factors of 2 on the right hand side of each equation are
due to equal contributions from the two valleys corresponding to every
patch. 

\begin{figure}
(a)\includegraphics[width=0.9\columnwidth]{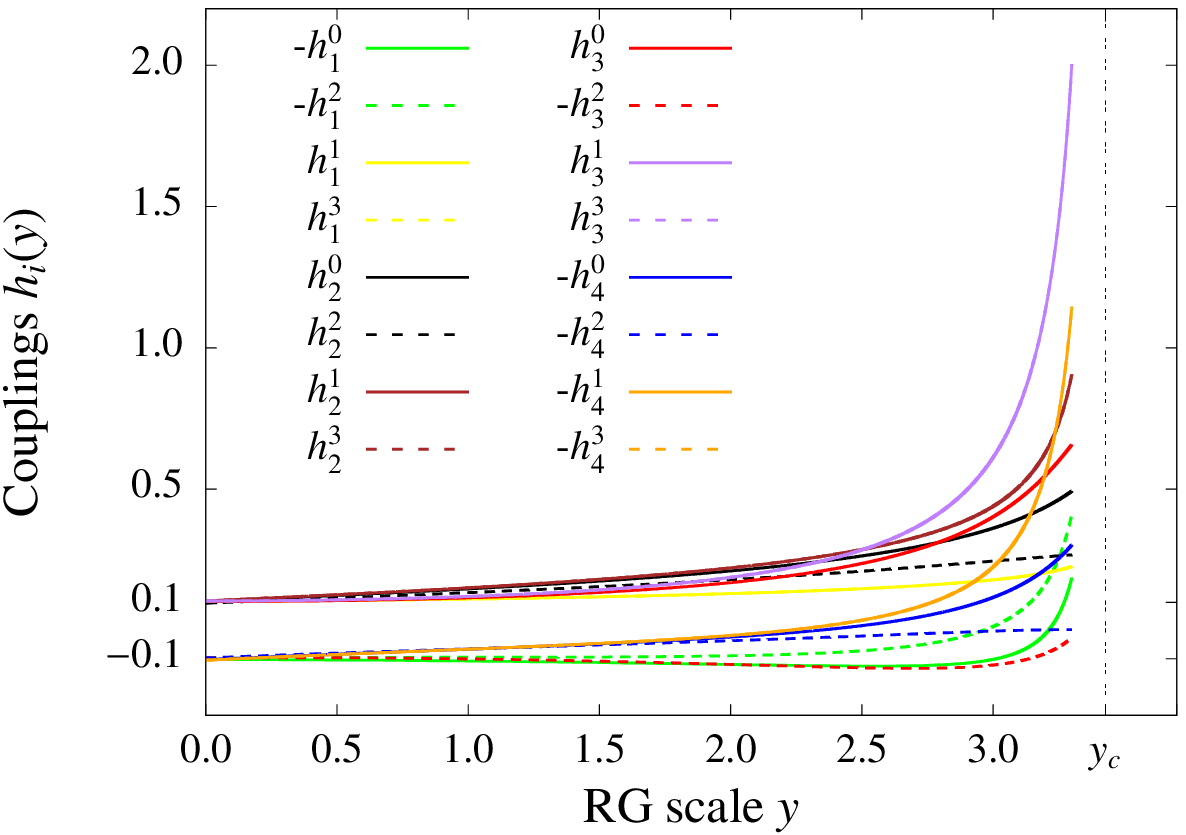}$\qquad\qquad$

(b)\includegraphics[width=0.9\columnwidth]{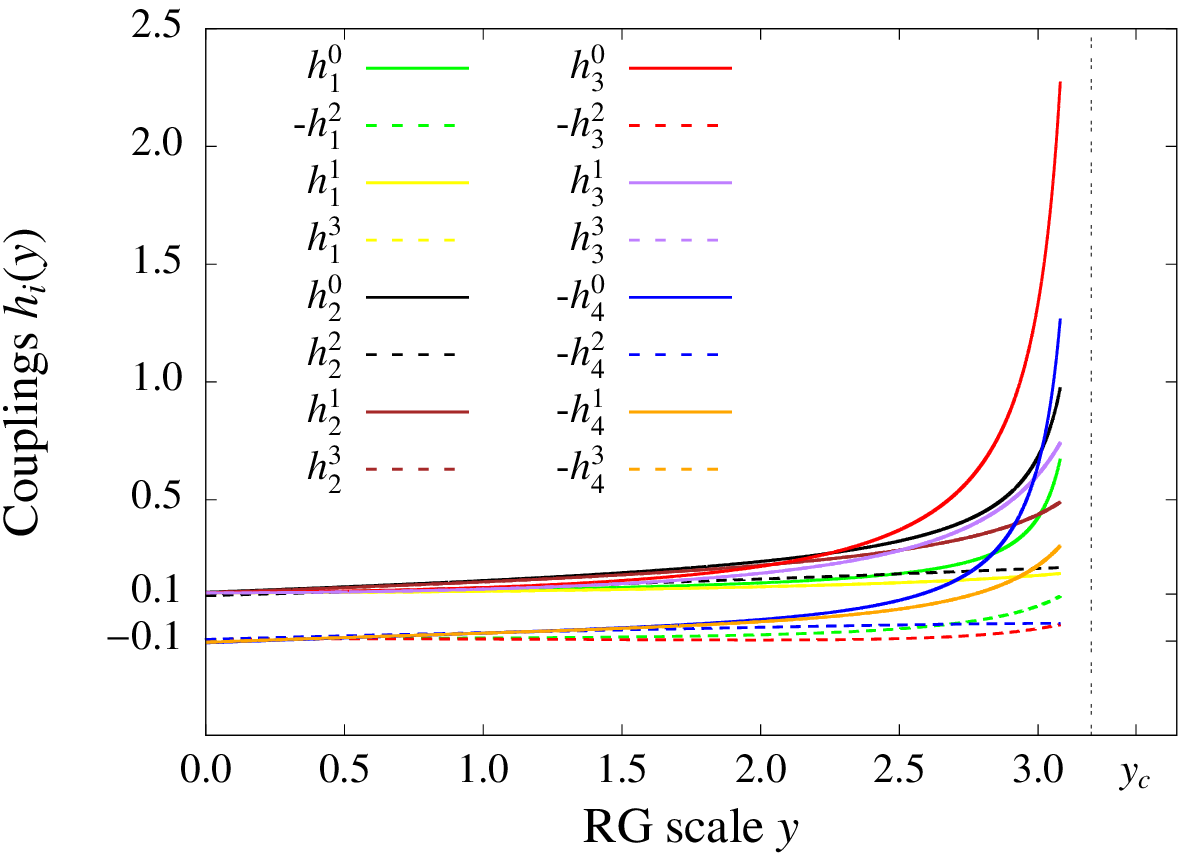}$\qquad\qquad$

\caption{\label{rg}The RG flows for (a) $M=4$ meV and (b) $M=9$ meV where
the critical value of spin-splitting $M_{c}\approx6.1$ meV. The fixed
point value $y_{c}\approx3.43$ for (a) and $y_{c}\approx3.2$ for
(b) above. Here, the initial values for each of the dimensionless
couplings is taken to be $0.1$, and a Hund's splitting of $5\%$
($\frac{|h_{i}^{\sigma\overline{\sigma}}-h_{i}^{\sigma\sigma}|}{|h_{i}^{\sigma\sigma}|}=0.05$)
is introduced initially such that $h_{i}^{\sigma\overline{\sigma}}>h_{i}^{\sigma\sigma}$
for $i=1-4$ and $\sigma=\uparrow,\downarrow$. Clearly, the leading
couplings near the instability threshold correspond to spin-antiparallel
configurations for $M<M_{c}$, while the spin $\uparrow$ component
of each of the couplings dominates for $M>M_{c}$. Here the couplings
$h_{i}^{\uparrow\uparrow}$, $h_{i}^{\downarrow\downarrow}$,$h_{i}^{\uparrow\downarrow}$and
$h_{i}^{\downarrow\uparrow}$ are denoted respectively by $h_{i}^{0}$,
$h_{i}^{2}$, $h_{i}^{1}$ and $h_{i}^{3}$, for clarity. The factors
of $v_{\uparrow}$ and $v_{\downarrow}$ have been absorbed into the
couplings constants $h_{i}^{\ell}$ ($\ell=0-3$) in the above plots,
for simplicity in notation.}
\end{figure}

\begin{figure}
(a)\includegraphics[width=0.9\columnwidth]{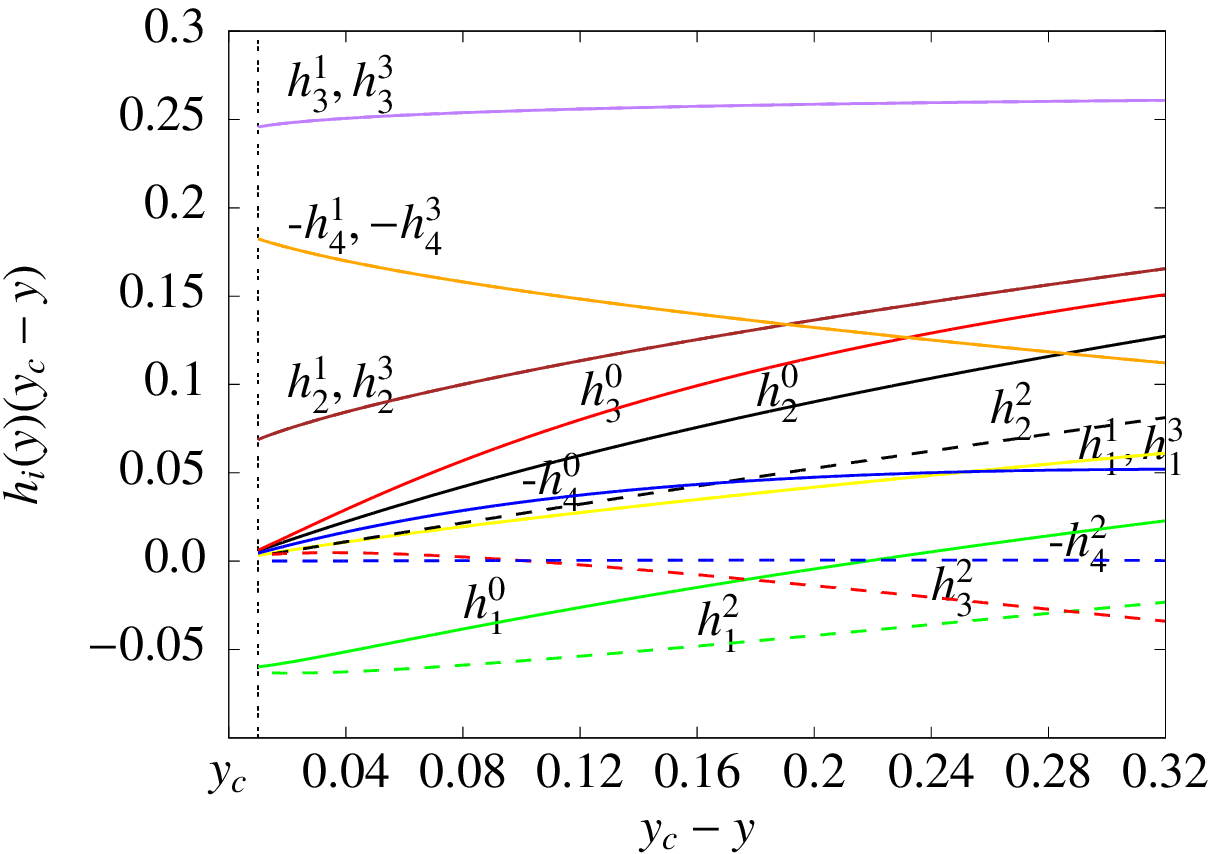}

(b)\includegraphics[width=0.9\columnwidth]{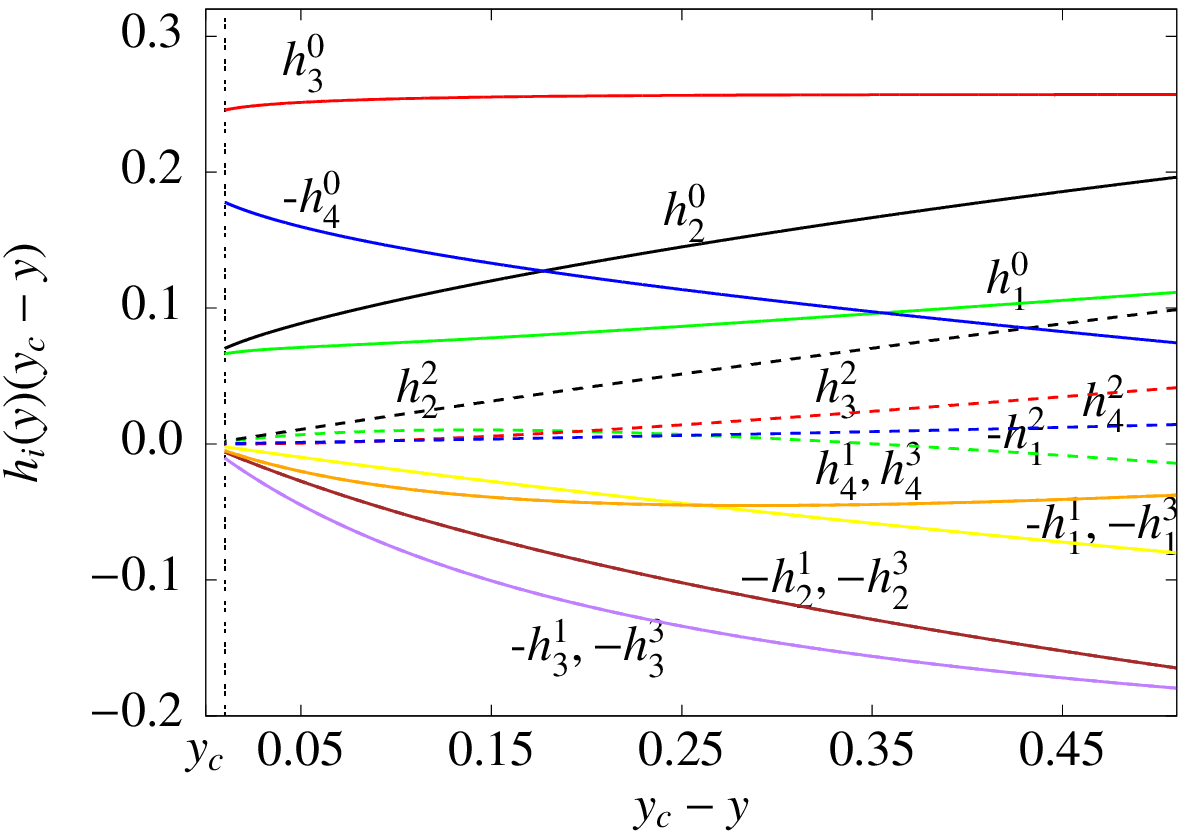}

\caption{\label{fig:fixedpoint}The order of the fixed-point values for the
different couplings $h_{i}^{\ell}$($\ell=0-3$) at the critical point
$y_{c}$ for (a) $M=4$ meV and (b) $M=9$ meV, where the critical
spin-splitting $M_{c}\approx6.1$ meV. Here, the initial values for
each of the dimensionless couplings is taken to be $0.1$, and a Hund's
splitting of $5\%$ ($\frac{|h_{i}^{\sigma\overline{\sigma}}-h_{i}^{\sigma\sigma}|}{|h_{i}^{\sigma\sigma}|}=0.05$)
is introduced initially such that $h_{i}^{\sigma\overline{\sigma}}>h_{i}^{\sigma\sigma}$
for $i=1-4$ and $\sigma=\uparrow,\downarrow$. The above plots show
the evolution of $h_{i}^{\ell}(y)$ as a function of $(y_{c}-y)$
close to the fixed point $y_{c}$, where each coupling constant $h_{i}^{\ell}$
has an asymptodic form $\frac{g_{i}^{\ell}}{y_{c}-y}$, and the $y$-intercepts
of curves shown give an estimate of the fixed-point values $g_{i}^{\ell}$
for the different couplings. This clearly indicates that the leading
couplings for $M<M_{c}$ correspond to the $\ell=1$
and $\ell=3$ channels, for spin-antiparallel configurations, in
the presence of a finite Hund's splitting, while for $M>M_{c}$ these
correspond to the spin $\uparrow$($\ell=0$) channel. As in Fig.\ref{rg},the
factors of $v_{\uparrow}$ and $v_{\downarrow}$ corresponding to each
coupling have been absorbed into $h_{i}^{\ell}$ ($\ell=0-3$). }
\end{figure}

\section*{4. \label{sec4}Susceptibilities}

In order to investigate the possible electronic instabilities in this
system, we shall now evaluate the susceptibilities $\chi$ for various
types of order, by introducing infinitesimal test vertices corresponding
to different kinds of pairing into the action, such as $\triangle_{a}\psi_{a\sigma}^{\dagger}\psi_{a\sigma^{\prime}}^{\dagger}+\triangle_{a}^{*}\psi_{a\sigma}\psi_{a\sigma^{\prime}}$
for the patch $a=1,2$ (where the spin labels $\sigma,\sigma^{\prime}$
on the fermions denote the presence or absence of the phase factors
$\exp[i\theta_{k}]$) corresponding to particle-particle pairing on
a patch\cite{nandkishore2012chiral,kundu2017role}. 

The renormalization of the test vertex corresponding to particle-particle
pairing on a patch is governed by the equation
\begin{equation}
\frac{\partial}{\partial y}\left(\begin{array}{c}
\Delta_{1}\\
\Delta_{2}
\end{array}\right)=2v_{\sigma}v_{\overline{\sigma}}\left(\begin{array}{cc}
h_{4}^{\sigma\overline{\sigma}} & h_{3}^{\sigma\overline{\sigma}}\\
h_{3}^{\sigma\overline{\sigma}} & h_{4}^{\sigma\overline{\sigma}}
\end{array}\right)\left(\begin{array}{c}
\Delta_{1}\\
\Delta_{2}
\end{array}\right)\label{eq:22}
\end{equation}
 since we can only consider Cooper pairing in the $p$-wave channel
for effectively spinless electrons on the TCI surface, where $\sigma,\overline{\sigma}=\uparrow,\downarrow$
in this case. By transforming to the eigenvector basis, we can obtain
different possible order parameters, and choose the one corresponding
to the most negative eigenvalue. The vertices with positive eigenvalues
are suppressed under RG flow. The renormalization equations for the
test vertices for other kinds of pairing can be similarly obtained.
The diagrams corresponding to the renormalization of the different
pairing vertices considered by us are shown in \cite{kundu2017role},
and we consider similar kinds of pairing here, although the total
number of instabilities possible increases in this case due to the
lifting of spin degeneracy by the Zeeman splitting term in the Hamiltonian. 

Each of the couplings has an asymptotic form $h_{i}(y)=\frac{g_{i}^{\sigma\sigma^{\prime}}}{y_{c}-y}$\cite{furukawa1998truncation,nandkishore2012chiral}
at the threshold. At an electronic instability, the most divergent
susceptibility $\chi$ determines the nature of the ordered phase.
The exponents $\alpha$ for susceptibilities $\chi$ corresponding
to the various order parameters (which have a general form $\chi\propto(y_{c}-y)^{\alpha}$)
are functions of the fixed point values of the couplings $g_{i}^{\sigma\sigma\prime}$.
The channel where the instability is most likely to take place has
the most singular susceptibility, i.e. the most negative value of
$\alpha$. By substituting the asymptodic form for the couplings $h_{i}^{\sigma\sigma^{\prime}}$
into the above eq.\ref{eq:22} and the corresponding equations for
other kinds of pairing, we have obtained the exponents $\alpha$ for
intrapatch $p$-wave pairing, charge-density wave, spin-density wave,
uniform spin, charge compressibility ($\kappa$) and finite-momentum
$\pi$ pairing, which are given as follows- 
\[
\alpha_{pwave}=2(g_{4}^{\sigma\overline{\sigma}}-g_{3}^{\sigma\overline{\sigma}})v_{\sigma}v_{\overline{\sigma}}
\]
\begin{align*}
\alpha_{\kappa} & =(g_{1}^{\sigma\sigma}(v_{\sigma})^{2}+g_{4}^{\sigma\overline{\sigma}}(v_{\sigma}v_{\overline{\sigma}})-g_{2}^{\sigma\sigma}(v_{\sigma})^{2}\\
 & -g_{2}^{\sigma\overline{\sigma}}(v_{\sigma}v_{\overline{\sigma}}))d_{2}(y_{c})
\end{align*}
\[
\alpha_{s}=-(g_{4}^{\sigma\overline{\sigma}}+g_{1}^{\sigma\overline{\sigma}})(v_{\sigma}v_{\overline{\sigma}})d_{2}(y_{c})
\]
\[
\alpha_{\pi}^{\sigma\overline{\sigma}}=(g_{1}^{\sigma\overline{\sigma}}+g_{2}^{\sigma\overline{\sigma}})(v_{\sigma}v_{\overline{\sigma}})d_{3}(y_{c})
\]
\[
\alpha_{\pi}^{\sigma\sigma}=(v_{\sigma})^{2}(g_{1}^{\sigma\sigma}+g_{2}^{\sigma\sigma})d_{3}(y_{c})
\]
\begin{alignat*}{1}
\alpha_{CDW} & =(g_{1}^{\sigma\sigma}(v_{\sigma})^{2}-g_{3}^{\sigma\overline{\sigma}}(v_{\sigma}v_{\overline{\sigma}})+g_{1}^{\sigma\overline{\sigma}}(v_{\sigma}v_{\overline{\sigma}})\\
 & -g_{2}^{\sigma\sigma}(v_{\sigma})^{2})d_{1}(y_{c})
\end{alignat*}
\begin{equation}
\alpha_{SDW}=-2(g_{3}^{\sigma\overline{\sigma}}+g_{2}^{\sigma\overline{\sigma}})(v_{\sigma}v_{\overline{\sigma}})d_{1}(y_{c})\label{eq:23}
\end{equation}
The above expressions correspond to various spin combinations $\sigma,\overline{\sigma}=\uparrow,\downarrow$.

\section*{5. \label{sec5}Ladder rg equations in the absence of hund's splitting
of interactions}

\begin{figure}
\includegraphics[width=1\columnwidth]{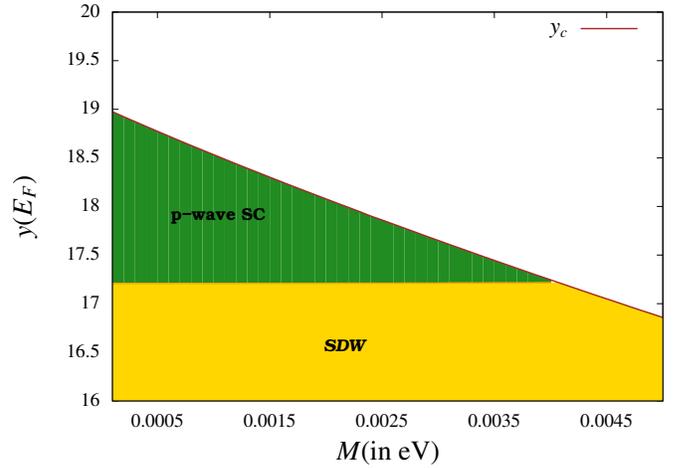}

\caption{\label{fig:The-phase-diagram}The phase diagram for y($E_{F}$) as
a function of the spin-splitting $M$ when the initial value of each
of the dimensionless RG couplings is chosen to be equal to $0.03$.
This shows that for large electron densities, it is possible to stabilize
p-wave superconductivity for a range of values of the Fermi energy
$E_{F}$, up till $M\sim4$ meV (in this case). The values on the
$y$-axis as well as the value of $M$ up to which $p$-wave superconductivity
may be stabilized depend on the initial interaction strength. The
latter decreases with an increase in the strength of electronic interactions. }
\end{figure}

Let us first consider a situation where the various components of
interactions $h_{i}$ in the different scattering channels $i=1-4$
are taken to be identical initially, with no Hund's splitting present.
In this case, we find that even for a very small value of Zeeman splitting
$M$, the leading components of the different kinds of interactions
near the fixed point $y_{c}$ correspond to spin $\uparrow$ (i.e.
the $\ell=0$ channel). Now, if we introduce test vertices for different
kinds of pairing and calculate the exponents for the divergence of
the respective susceptibilities, we find that each of the exponents
$\alpha$ is either positive or numerically close to zero. This indicates
the absence of any electronic instabilities in this case. Clearly,
$p$-wave superconductivity cannot be stabilized at energies corresponding
to the fixed point of the parquet RG. However, when the Fermi energy
$E_{F}$ associated with the system exceeds the energy $\omega_{c}$
corresponding to the critical point $y_{c}$, the RG flow must be
terminated at $E_{F}$, and any possible instabilities will then depend
on the order of the different couplings at the Fermi energy. These
are determined using a ladder RG approach, which is described in \cite{chubukov2009renormalization}. 

Two kinds of vertices continue to flow logarithmically at energies
below the Fermi energy $E_{F}$: vertices with zero total momentum,
and with total momentum exactly equal to the nesting vector $Q$ in
two dimensions. The vertices with zero total momentum are the $h_{3}$
and $h_{4}$ terms in our RG analysis and the vertices with total
momentum Q are the $h_{1}$, $h_{2}$ and $h_{3}$ terms. The values
of $h_{i}$ at $E_{F}$ act as the bare couplings for the theory at
$E<E_{F}$. There are two kinds of $h_{3}$ vertices with a momentum
transfer $Q$, $h_{3a}$and $h_{3b}$(for a detailed discussion,
please refer to \cite{chubukov2009renormalization}) and we denote
the $h_{3}$ vertex with zero total momentum as $h_{3c}$. Following
\cite{chubukov2009renormalization}, we shall refer to the vertices
with zero total momentum as $h_{i}(0)$ and the vertices with total
momentum $Q$ as $h_{i}(Q)$. The ladder RG equations are obtained
by considering those diagrams which still yield a logarithmic divergence
\cite{chubukov2009renormalization}. 

The ladder RG equations for our system, where now $y\equiv\ln[\frac{E_{F}}{\omega}]$,
are given as follows-
\begin{eqnarray*}
\frac{dh_{1}^{\sigma\sigma}(Q)}{dy} & = & 2((2h_{3a}^{\sigma\sigma}(Q)h_{3b}^{\sigma\sigma}(Q)-(h_{3a}^{\sigma\sigma}(Q))^{2})v_{\sigma}^{2}\\
 &  & -(h_{1}^{\sigma\sigma}(Q))^{2}v_{\sigma}^{2}-(h_{3a}^{\sigma\overline{\sigma}}(Q))(h_{3a}^{\overline{\sigma}\sigma}(Q))v_{\overline{\sigma}}^{2}-\\
 &  & (h_{1}^{\sigma\overline{\sigma}}(Q))(h_{1}^{\overline{\sigma}\sigma}(Q))v_{\overline{\sigma}}^{2}+2(h_{1}^{\sigma\sigma}(Q))(h_{2}^{\sigma\sigma}(Q))v_{\sigma}^{2})
\end{eqnarray*}
\begin{eqnarray*}
\frac{dh_{1}^{\overline{\sigma}\overline{\sigma}}(Q)}{dy} & = & 2((2(h_{3a}^{\overline{\sigma}\overline{\sigma}}(Q))(h_{3b}^{\overline{\sigma}\overline{\sigma}}(Q))-(h_{3a}^{\overline{\sigma}\overline{\sigma}}(Q))^{2})v_{\overline{\sigma}}^{2}\\
 &  & -(h_{1}^{\overline{\sigma}\overline{\sigma}}(Q))^{2}v_{\overline{\sigma}}^{2}-(h_{3a}^{\overline{\sigma}\sigma}(Q))(h_{3a}^{\sigma\overline{\sigma}}(Q))v_{\sigma}^{2}-\\
 &  & (h_{1}^{\overline{\sigma}\sigma}(Q))(h_{1}^{\sigma\overline{\sigma}}(Q))v_{\sigma}^{2}+\\
 &  & 2(h_{1}^{\overline{\sigma}\overline{\sigma}}(Q))(h_{2}^{\overline{\sigma}\overline{\sigma}}(Q))v_{\overline{\sigma}}^{2})\qquad
\end{eqnarray*}
\begin{eqnarray*}
\frac{dh_{1}^{\sigma\overline{\sigma}}(Q)}{dy} & = & 2(h_{3a}^{\sigma\overline{\sigma}}(Q)(h_{3b}^{\overline{\sigma}\overline{\sigma}}(Q)-h_{3a}^{\overline{\sigma}\overline{\sigma}}(Q))v_{\overline{\sigma}}^{2}\\
 &  & +h_{3a}^{\sigma\overline{\sigma}}(Q)(h_{3b}^{\sigma\sigma}(Q)-h_{3a}^{\sigma\sigma}(Q))v_{\sigma}^{2}+h_{1}^{\sigma\overline{\sigma}}(Q)\\
 &  & (-(h_{1}^{\overline{\sigma}\overline{\sigma}}(Q))v_{\overline{\sigma}}^{2}-(h_{1}^{\sigma\sigma}(Q))v_{\sigma}^{2}+(h_{2}^{\sigma\sigma}(Q))v_{\sigma}^{2}\\
 &  & +(h_{2}^{\overline{\sigma}\overline{\sigma}}(Q))v_{\overline{\sigma}}^{2}))
\end{eqnarray*}
\begin{eqnarray*}
\frac{dh_{1}^{\overline{\sigma}\sigma}(Q)}{dy} & = & 2(h_{3a}^{\overline{\sigma}\sigma}(Q)(h_{3b}^{\overline{\sigma}\overline{\sigma}}(Q)-h_{3a}^{\overline{\sigma}\overline{\sigma}}(Q))v_{\overline{\sigma}}^{2}\\
 &  & +h_{3a}^{\overline{\sigma}\sigma}(Q)(h_{3b}^{\sigma\sigma}(Q)-h_{3a}^{\sigma\sigma}(Q))v_{\sigma}^{2}+h_{1}^{\overline{\sigma}\sigma}(Q)\\
 &  & (-(h_{1}^{\overline{\sigma}\overline{\sigma}}(Q))v_{\overline{\sigma}}^{2}-(h_{1}^{\sigma\sigma}(Q))v_{\sigma}^{2}+(h_{2}^{\sigma\sigma}(Q))v_{\sigma}^{2}\\
 &  & +(h_{2}^{\overline{\sigma}\overline{\sigma}}(Q))v_{\overline{\sigma}}^{2}))
\end{eqnarray*}
\[
\frac{dh_{2}^{\sigma\sigma}(Q)}{dy}=2((h_{2}^{\sigma\sigma}(Q))^{2}+(h_{3b}^{\sigma\sigma}(Q))^{2})v_{\sigma}^{2}\qquad\qquad\qquad
\]
\[
\frac{dh_{2}^{\overline{\sigma}\overline{\sigma}}(Q)}{dy}=2((h_{2}^{\overline{\sigma}\overline{\sigma}}(Q))^{2}+(h_{3b}^{\overline{\sigma}\overline{\sigma}}(Q))^{2})v_{\overline{\sigma}}^{2}\qquad\qquad\qquad
\]
\[
\frac{dh_{2}^{\sigma\overline{\sigma}}(Q)}{dy}=2((h_{2}^{\sigma\overline{\sigma}}(Q))^{2}+(h_{3b}^{\sigma\overline{\sigma}}(Q))^{2})v_{\sigma}v_{\overline{\sigma}}\qquad\qquad\qquad
\]
\[
\frac{dh_{2}^{\overline{\sigma}\sigma}(Q)}{dy}=2((h_{2}^{\overline{\sigma}\sigma}(Q))^{2}+(h_{3b}^{\overline{\sigma}\sigma}(Q))^{2})v_{\sigma}v_{\overline{\sigma}}\qquad\qquad\qquad
\]
\begin{eqnarray*}
\frac{dh_{3a}^{\sigma\sigma}(Q)}{dy} & = & 2(2h_{1}^{\sigma\sigma}(Q)(h_{3b}^{\sigma\sigma}(Q)\\
 &  & -h_{3a}^{\sigma\sigma}(Q))v_{\sigma}^{2}+2h_{2}^{\sigma\sigma}(Q)h_{3a}^{\sigma\sigma}(Q)v_{\sigma}^{2}\\
 &  & -h_{1}^{\overline{\sigma}\sigma}(Q)h_{3a}^{\sigma\overline{\sigma}}(Q)v_{\overline{\sigma}}^{2}-h_{1}^{\sigma\overline{\sigma}}(Q)h_{3a}^{\overline{\sigma}\sigma}(Q)v_{\overline{\sigma}}^{2})\qquad
\end{eqnarray*}
\[
\frac{dh_{3b}^{\sigma\sigma}(Q)}{dy}=4h_{2}^{\sigma\sigma}(Q)h_{3b}^{\sigma\sigma}(Q)v_{\sigma}^{2}\qquad\qquad\qquad\qquad\qquad\qquad
\]
\[
\frac{dh_{3c}^{\sigma\sigma}(0)}{dy}=-4h_{4}^{\sigma\sigma}(0)h_{3c}^{\sigma\sigma}(0)v_{\sigma}^{2}\qquad\qquad\qquad\qquad\qquad\qquad
\]
\begin{eqnarray*}
\frac{dh_{3a}^{\overline{\sigma}\overline{\sigma}}(Q)}{dy} & = & 2(2h_{1}^{\overline{\sigma}\overline{\sigma}}(Q)(h_{3b}^{\overline{\sigma}\overline{\sigma}}(Q)\\
 &  & -h_{3a}^{\overline{\sigma}\overline{\sigma}}(Q))v_{\overline{\sigma}}^{2}+2h_{2}^{\overline{\sigma}\overline{\sigma}}(Q)h_{3a}^{\overline{\sigma}\overline{\sigma}}(Q)v_{\overline{\sigma}}^{2}\\
 &  & -h_{1}^{\overline{\sigma}\sigma}(Q)h_{3a}^{\sigma\overline{\sigma}}(Q)v_{\sigma}^{2}-h_{1}^{\sigma\overline{\sigma}}(Q)h_{3a}^{\overline{\sigma}\sigma}(Q)v_{\sigma}^{2})\qquad
\end{eqnarray*}
\[
\frac{dh_{3b}^{\overline{\sigma}\overline{\sigma}}(Q)}{dy}=4h_{2}^{\overline{\sigma}\overline{\sigma}}(Q)h_{3b}^{\overline{\sigma}\overline{\sigma}}(Q)v_{\overline{\sigma}}^{2}\qquad\qquad\qquad\qquad\qquad\qquad
\]
\[
\frac{dh_{3c}^{\overline{\sigma}\overline{\sigma}}(0)}{dy}=-4h_{4}^{\overline{\sigma}\overline{\sigma}}(0)h_{3c}^{\overline{\sigma}\overline{\sigma}}(0)v_{\overline{\sigma}}^{2}\qquad\qquad\qquad\qquad\qquad\qquad
\]
\begin{eqnarray*}
\frac{dh_{3a}^{\sigma\overline{\sigma}}(Q)}{dy} & = & 2(h_{1}^{\sigma\overline{\sigma}}(Q)(h_{3b}^{\sigma\sigma}(Q)-h_{3a}^{\sigma\sigma}(Q))v_{\sigma}^{2}\\
 &  & +h_{1}^{\sigma\overline{\sigma}}(Q)(h_{3b}^{\overline{\sigma}\overline{\sigma}}(Q)-h_{3a}^{\overline{\sigma}\overline{\sigma}}(Q))v_{\overline{\sigma}}^{2}+\\
 &  & h_{3a}^{\sigma\overline{\sigma}}(Q)(h_{2}^{\sigma\sigma}(Q)v_{\sigma}^{2}+h_{2}^{\overline{\sigma}\overline{\sigma}}(Q)v_{\overline{\sigma}}^{2}\\
 &  & -h_{1}^{\sigma\sigma}(Q)v_{\sigma}^{2}-h_{1}^{\overline{\sigma}\overline{\sigma}}(Q)v_{\overline{\sigma}}^{2}))\qquad
\end{eqnarray*}
\[
\frac{dh_{3b}^{\sigma\overline{\sigma}}(Q)}{dy}=4h_{2}^{\sigma\overline{\sigma}}(Q)h_{3b}^{\sigma\overline{\sigma}}(Q)v_{\sigma}v_{\overline{\sigma}}\qquad\qquad\qquad
\]
\[
\frac{dh_{3c}^{\sigma\overline{\sigma}}(0)}{dy}=-4h_{4}^{\sigma\overline{\sigma}}(0)h_{3c}^{\sigma\overline{\sigma}}(0)v_{\sigma}v_{\overline{\sigma}}\qquad\qquad\qquad
\]
\\
\begin{eqnarray*}
\frac{dh_{3a}^{\overline{\sigma}\sigma}(Q)}{dy} & = & 2(h_{1}^{\overline{\sigma}\sigma}(Q)(h_{3b}^{\sigma\sigma}(Q)-h_{3a}^{\sigma\sigma}(Q))v_{\sigma}^{2}\\
 &  & +h_{1}^{\overline{\sigma}\sigma}(Q)(h_{3b}^{\overline{\sigma}\overline{\sigma}}(Q)-h_{3a}^{\overline{\sigma}\overline{\sigma}}(Q))v_{\overline{\sigma}}^{2}+\\
 &  & h_{3a}^{\overline{\sigma}\sigma}(Q)(h_{2}^{\sigma\sigma}(Q)v_{\sigma}^{2}+h_{2}^{\overline{\sigma}\overline{\sigma}}(Q)v_{\overline{\sigma}}^{2}\\
 &  & -h_{1}^{\sigma\sigma}(Q)v_{\sigma}^{2}-h_{1}^{\overline{\sigma}\overline{\sigma}}(Q)v_{\overline{\sigma}}^{2}))\qquad
\end{eqnarray*}
\[
\frac{dh_{3b}^{\overline{\sigma}\sigma}(Q)}{dy}=4h_{2}^{\overline{\sigma}\sigma}(Q)h_{3b}^{\overline{\sigma}\sigma}(Q)v_{\sigma}v_{\overline{\sigma}}\qquad\qquad\qquad
\]
\[
\frac{dh_{3c}^{\overline{\sigma}\sigma}(0)}{dy}=-4h_{4}^{\overline{\sigma}\sigma}(0)h_{3c}^{\overline{\sigma}\sigma}(0)v_{\sigma}v_{\overline{\sigma}}\qquad\qquad\qquad
\]
\[
\frac{dh_{4}^{\sigma\sigma}(0)}{dy}=-2((h_{3c}^{\sigma\sigma}(0))^{2}+(h_{4}^{\sigma\sigma}(0))^{2})v_{\sigma}^{2}\qquad
\]
\[
\frac{dh_{4}^{\overline{\sigma}\overline{\sigma}}(0)}{dy}=-2((h_{3c}^{\overline{\sigma}\overline{\sigma}}(0))^{2}+(h_{4}^{\overline{\sigma}\overline{\sigma}}(0))^{2})v_{\overline{\sigma}}^{2}\qquad
\]
\[
\frac{dh_{4}^{\sigma\overline{\sigma}}(0)}{dy}=-2((h_{3c}^{\sigma\overline{\sigma}}(0))^{2}+(h_{4}^{\sigma\overline{\sigma}}(0))^{2})v_{\sigma}v_{\overline{\sigma}}\qquad
\]
\begin{equation}
\frac{dh_{4}^{\overline{\sigma}\sigma}(0)}{dy}=-2((h_{3c}^{\overline{\sigma}\sigma}(0))^{2}+(h_{4}^{\overline{\sigma}\sigma}(0))^{2})v_{\sigma}v_{\overline{\sigma}}\label{eq:24}
\end{equation}

From the above equations, we find 
\[
\frac{d(h_{3c}^{\sigma\overline{\sigma}}(0)-h_{4}^{\sigma\overline{\sigma}}(0))}{dy}=2(h_{3c}^{\sigma\overline{\sigma}}(0)-h_{4}^{\sigma\overline{\sigma}}(0))^{2}v_{\sigma}v_{\overline{\sigma}}
\]
 for the superconducting vertex. These equations can be solved to
give 
\[
h_{3}^{\sigma\overline{\sigma}}(0)-h_{4}^{\sigma\overline{\sigma}}(0)=\frac{(h_{3}^{\sigma\overline{\sigma}})_{E_{F}}-(h_{4}^{\sigma\overline{\sigma}})_{E_{F}}}{1-2v_{\sigma}v_{\overline{\sigma}}((h_{3}^{\sigma\overline{\sigma}})_{E_{F}}-(h_{4}^{\sigma\overline{\sigma}})_{E_{F}})\log[\frac{E_{F}}{\omega}]}
\]
A similar situation arises for the SDW instability in this regime.
The competition between these instabilities depends on the respective
energies at which different combinations of couplings diverge, and
thus, on their values at the Fermi energy $E_{F}$. The first instability
occurs in the channel for which the coupling at $\omega\sim E_{F}$
is the largest.

Thus, we find that for relatively large electron densities, when the
Fermi energy $E_{F}$ exceeds the energy ($\omega_{c}$) corresponding
to the critical point of the RG flow $y_{c}$, a $p$-wave superconducting
order can be stabilized on the TCI surface up to a small value of
the spin-splitting $M$ ($\sim1$ meV). For larger values of Zeeman
splitting introduced by an external magnetization, we find that a
spin density wave (SDW) modulation may be possible over and above
the expected uniform spin polarization on the surface, if the number
density of electrons is sufficiently large. Although $p$-wave superconductivity
is degraded even by a very small value of external magnetization in the
absence of Hund's splitting, it is thus possible to stabilize this
phase over a range of electron densities (and corresponding Fermi
energies $E_{F}$). A phase diagram for $y(E_{F}$) as a function
of the spin-splitting term $M$ is shown in Fig.\ref{fig:The-phase-diagram}
for an initial value of $0.03$ for each of the dimensionless couplings.
It should be noted that the exact values on the $y-$axis, as well
as the value of spin-splitting $M$ (on the $x$-axis) beyond which
$p$-wave superconductivity is no longer possible, is dependent
on the initial interaction strength being considered. In particular,
we find that the range of values of $M$ for which $p$-wave superconductivity
may be stabilized decreases with an increase in the strength of electronic
interactions.

\section*{6.\label{sec6} Critical values of spin-splitting for a finite hund's
interaction}

\begin{figure}
\includegraphics[width=1\columnwidth]{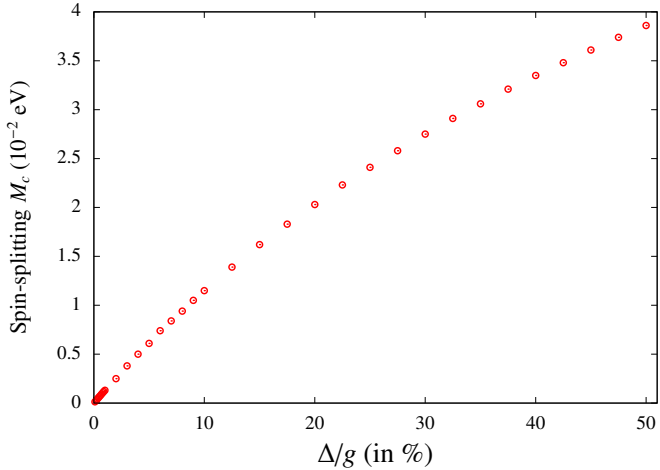}

\caption{\label{fig:mvsg}The behavior of the critical value of the spin-splitting
$M_{c}$(in eV) as a function of the Hund's splitting $\Delta$ as
a percentage of the initial interaction $g$ ( i.e. $h_{i}^{\sigma\sigma}=g$
initially for $i=1-4$ and $\sigma=\uparrow,\downarrow$), where $g=0.1$
in this case. We find this behavior to be extremely insensitive to
the initial value of interactions.}
\end{figure}

For a multiorbital system like Pb$_{1-x}$Sn$_{x}$Te, one must also
take into account the effects of Hund's splitting. This effect can
be built into our RG analysis by assuming the initial values of interactions
in each of the scattering channels $i$ to be such that $(h_{i}^{\sigma\overline{\sigma}}-h_{i}^{\sigma\sigma})>0$
(where $\sigma=\uparrow,\downarrow$). In \cite{kundu2017role}, we
have shown that $p$-wave superconductivity is favored on the surface
of Pb$_{1-x}$Sn$_{x}$Te even in the absence of Hund's splitting
(i.e. when the interactions in the different channels are chosen to
be identical initially). As seen in the previous section, in the presence
of an external magnetization, $p$-wave superconductivity is destroyed
(at the parquet level) even by a small value of Zeeman splitting.
However, this is no longer true if a finite Hund's splitting is introduced
initially. For a Hund's splitting of $\triangle$$=h_{i}^{\sigma\overline{\sigma}}-h_{i}^{\sigma\sigma}$
(for each scattering channel $i$, where $\sigma=\uparrow,\downarrow$),
$p$-wave superconductivity continues to be the leading instability
at the parquet level up to a finite value of the Zeeman splitting
$M$ (which depends on the value of $\triangle$ being considered).
Corresponding to each value of $\Delta$, a critical value of the
spin-splitting $M_{c}$ is obtained such that for $M>M_{c}$, $p$-wave
superconductivity is no longer possible. The variation of $M_{c}$
as a function of the percentage Hund's splitting $\frac{\Delta}{g}$
(where $g$ denotes the initial value chosen for $h_{i}^{\sigma\sigma}$
for $i=1-4$ with $\sigma=\uparrow,\downarrow$) is shown in Fig.\ref{fig:mvsg}
for $g=0.1$. The behavior of $M_{c}$ as a function of $\frac{\Delta}{g}$
turns out to be remarkably insensitive to value of $g$, i.e. the
initial interaction strength being considered (within the regime where
perturbation theory is valid). To illustrate the nature of the most
divergent couplings in the two limits, RG flows for $M<M_{c}$ and
$M>M_{c}$ with a dimensionless initial repulsive interaction of $0.1$
and a Hund's splitting of $5\%$ ($\frac{|h_{i}^{\sigma\overline{\sigma}}-h_{i}^{\sigma\sigma}|}{|h_{i}^{\sigma\sigma}|}=0.05$)
introduced initially, where the critical value of the Zeeman splitting
$M_{c}\approx6.1$ meV, are shown in the Fig \ref{rg}. The corresponding
behavior of $h_{i}^{\sigma\sigma^{\prime}}(y)(y_{c}-y)$ as a function
of $(y_{c}-y)$, which illustrates the order of the fixed point values
$g_{i}^{\sigma\sigma^{\prime}}$ for the different couplings in the
above-mentioned two cases, is shown in the Fig. \ref{fig:fixedpoint}.

\section*{7.\label{sec:7.-sec7} Summary and discussion }

In conclusion, we have studied the effect of an external magnetization
on chiral $p$-wave superconductivity (predicted by us in \cite{kundu2017role})
on the (001) surface of the multiorbital crystalline topological insulator
Pb$_{1-x}$Sn$_{x}$Te, which was found to be sensitive to the sign
of the Hund's splitting. We have shown that in the absence of Hund's
splitting of interactions, the $p$-wave superconductivity may be
destroyed even for very small values of the external magnetization.
However, robust $p$-wave superconductivity, stable against moderately
large values of the magnetization is obtained upon introduction of
a finite Hund's splitting of interactions, such that electrons with
spin-antiparallel configurations interact more strongly than those
with spin-parallel configurations. 

It should be kept in mind that the conclusions drawn from the perturbative
RG analysis are valid as long as the spin-splitting does not exceed
the characteristic energy scale at the critical point. Otherwise,
it would not be possible for interaction effects to dictate the ground
state properties since their characteristic energies would then fall
short of the spin-splitting scale. 

While we have studied the effects of a time-reversal symmetry breaking
perturbation on the surface superconductivity, further work is needed
to understand the effects of disorder, which in other unconventional
superconductors, is known to have a strong effect on their properties.
Given that the superconductivity in our case arises from Berry phase
effects, and not from Fermi surface deformations, we believe (see
ref. \cite{kundu2017role} for a discussion) that moderate amounts
of potential disorder will not cause destruction of the $p$-wave
superconducting order\cite{PhysRevLett.109.187003,nagai2015robust}.
This is unlike the case of Sr$_{2}$RuO$_{4}$, where $p$-wave superconductivity
is associated with Fermi surface deformations, and consequently, is
very sensitive to potential disorder \cite{PhysRevLett.80.161}. 

Recently, there have been reports of surface superconductivity induced
on the surface of Pb$_{0.6}$Sn$_{0.4}$Te by forming a mesoscopic
point contact using a nonsuperconducting metal \cite{das2016unexpected},
with a transition temperature in the range 3.7-6.5 K, although the
symmetry of the superconducting order was not confirmed. Our predictions
may be verified by examining the sensitivity of the superconducting
order to magnetic doping on the (001) surface of the TCI. 

Another interesting direction would be to further study the properties
of the $p$-wave superconductor in the presence of an external magnetization.
Given that proximity-induced chiral superconductivity recently led
to one-dimensional Majorana fermion modes in the hybrid system of
a magnetic topological insulator thin film coupled to a superconductor
\cite{he2017chiral}, a relevant question to address in this context
might be the coexistence of chiral $p$-wave superconductivity (which
is intrinsic in our case) with a Quantum Anomalous Hall state, induced
by the external magnetization, on the surface of Pb$_{1-x}$Sn$_{x}$Te.
\begin{acknowledgments}
SK acknowledges Debjyoti Burdhan for his help with some of the figures.
VT acknowledges DST for a Swarnajayanti grant (No. DST/SJF/PSA-0212012-13). 
\end{acknowledgments}

\bibliographystyle{apsrev4-1}

\begin{thebibliography}{33}%
\makeatletter
\providecommand \@ifxundefined [1]{%
 \@ifx{#1\undefined}
}%
\providecommand \@ifnum [1]{%
 \ifnum #1\expandafter \@firstoftwo
 \else \expandafter \@secondoftwo
 \fi
}%
\providecommand \@ifx [1]{%
 \ifx #1\expandafter \@firstoftwo
 \else \expandafter \@secondoftwo
 \fi
}%
\providecommand \natexlab [1]{#1}%
\providecommand \enquote  [1]{``#1''}%
\providecommand \bibnamefont  [1]{#1}%
\providecommand \bibfnamefont [1]{#1}%
\providecommand \citenamefont [1]{#1}%
\providecommand \href@noop [0]{\@secondoftwo}%
\providecommand \href [0]{\begingroup \@sanitize@url \@href}%
\providecommand \@href[1]{\@@startlink{#1}\@@href}%
\providecommand \@@href[1]{\endgroup#1\@@endlink}%
\providecommand \@sanitize@url [0]{\catcode `\\12\catcode `\$12\catcode
  `\&12\catcode `\#12\catcode `\^12\catcode `\_12\catcode `\%12\relax}%
\providecommand \@@startlink[1]{}%
\providecommand \@@endlink[0]{}%
\providecommand \url  [0]{\begingroup\@sanitize@url \@url }%
\providecommand \@url [1]{\endgroup\@href {#1}{\urlprefix }}%
\providecommand \urlprefix  [0]{URL }%
\providecommand \Eprint [0]{\href }%
\providecommand \doibase [0]{http://dx.doi.org/}%
\providecommand \selectlanguage [0]{\@gobble}%
\providecommand \bibinfo  [0]{\@secondoftwo}%
\providecommand \bibfield  [0]{\@secondoftwo}%
\providecommand \translation [1]{[#1]}%
\providecommand \BibitemOpen [0]{}%
\providecommand \bibitemStop [0]{}%
\providecommand \bibitemNoStop [0]{.\EOS\space}%
\providecommand \EOS [0]{\spacefactor3000\relax}%
\providecommand \BibitemShut  [1]{\csname bibitem#1\endcsname}%
\let\auto@bib@innerbib\@empty
\bibitem [{\citenamefont {Fu}(2011)}]{fu2011topological}%
  \BibitemOpen
  \bibfield  {author} {\bibinfo {author} {\bibfnamefont {L.}~\bibnamefont
  {Fu}},\ }\href@noop {} {\bibfield  {journal} {\bibinfo  {journal} {Phys. Rev.
  Lett.}\ }\textbf {\bibinfo {volume} {106}},\ \bibinfo {pages} {106802}
  (\bibinfo {year} {2011})}\BibitemShut {NoStop}%
\bibitem [{\citenamefont {Dziawa}\ \emph {et~al.}(2012)\citenamefont {Dziawa},
  \citenamefont {Kowalski}, \citenamefont {Dybko}, \citenamefont {Buczko},
  \citenamefont {Szczerbakow}, \citenamefont {Szot}, \citenamefont
  {{\L}usakowska}, \citenamefont {Balasubramanian}, \citenamefont {Wojek},
  \citenamefont {Berntsen} \emph {et~al.}}]{dziawa2012topological}%
  \BibitemOpen
  \bibfield  {author} {\bibinfo {author} {\bibfnamefont {P.}~\bibnamefont
  {Dziawa}}, \bibinfo {author} {\bibfnamefont {B.}~\bibnamefont {Kowalski}},
  \bibinfo {author} {\bibfnamefont {K.}~\bibnamefont {Dybko}}, \bibinfo
  {author} {\bibfnamefont {R.}~\bibnamefont {Buczko}}, \bibinfo {author}
  {\bibfnamefont {A.}~\bibnamefont {Szczerbakow}}, \bibinfo {author}
  {\bibfnamefont {M.}~\bibnamefont {Szot}}, \bibinfo {author} {\bibfnamefont
  {E.}~\bibnamefont {{\L}usakowska}}, \bibinfo {author} {\bibfnamefont
  {T.}~\bibnamefont {Balasubramanian}}, \bibinfo {author} {\bibfnamefont
  {B.~M.}\ \bibnamefont {Wojek}}, \bibinfo {author} {\bibfnamefont
  {M.}~\bibnamefont {Berntsen}},  \emph {et~al.},\ }\href@noop {} {\bibfield
  {journal} {\bibinfo  {journal} {Nat. Mater.}\ }\textbf {\bibinfo {volume}
  {11}},\ \bibinfo {pages} {1023} (\bibinfo {year} {2012})}\BibitemShut
  {NoStop}%
\bibitem [{\citenamefont {Hsieh}\ \emph {et~al.}(2012)\citenamefont {Hsieh},
  \citenamefont {Lin}, \citenamefont {Liu}, \citenamefont {Duan}, \citenamefont
  {Bansil},\ and\ \citenamefont {Fu}}]{hsieh2012topological}%
  \BibitemOpen
  \bibfield  {author} {\bibinfo {author} {\bibfnamefont {T.~H.}\ \bibnamefont
  {Hsieh}}, \bibinfo {author} {\bibfnamefont {H.}~\bibnamefont {Lin}}, \bibinfo
  {author} {\bibfnamefont {J.}~\bibnamefont {Liu}}, \bibinfo {author}
  {\bibfnamefont {W.}~\bibnamefont {Duan}}, \bibinfo {author} {\bibfnamefont
  {A.}~\bibnamefont {Bansil}}, \ and\ \bibinfo {author} {\bibfnamefont
  {L.}~\bibnamefont {Fu}},\ }\href@noop {} {\bibfield  {journal} {\bibinfo
  {journal} {Nat. Commun.}\ }\textbf {\bibinfo {volume} {3}},\ \bibinfo {pages}
  {982} (\bibinfo {year} {2012})}\BibitemShut {NoStop}%
\bibitem [{\citenamefont {Tanaka}\ \emph {et~al.}(2012)\citenamefont {Tanaka},
  \citenamefont {Ren}, \citenamefont {Sato}, \citenamefont {Nakayama},
  \citenamefont {Souma}, \citenamefont {Takahashi}, \citenamefont {Segawa},\
  and\ \citenamefont {Ando}}]{tanaka2012experimental}%
  \BibitemOpen
  \bibfield  {author} {\bibinfo {author} {\bibfnamefont {Y.}~\bibnamefont
  {Tanaka}}, \bibinfo {author} {\bibfnamefont {Z.}~\bibnamefont {Ren}},
  \bibinfo {author} {\bibfnamefont {T.}~\bibnamefont {Sato}}, \bibinfo {author}
  {\bibfnamefont {K.}~\bibnamefont {Nakayama}}, \bibinfo {author}
  {\bibfnamefont {S.}~\bibnamefont {Souma}}, \bibinfo {author} {\bibfnamefont
  {T.}~\bibnamefont {Takahashi}}, \bibinfo {author} {\bibfnamefont
  {K.}~\bibnamefont {Segawa}}, \ and\ \bibinfo {author} {\bibfnamefont
  {Y.}~\bibnamefont {Ando}},\ }\href@noop {} {\bibfield  {journal} {\bibinfo
  {journal} {Nature Phys.}\ }\textbf {\bibinfo {volume} {8}},\ \bibinfo {pages}
  {800} (\bibinfo {year} {2012})}\BibitemShut {NoStop}%
\bibitem [{\citenamefont {Xu}\ \emph {et~al.}(2012{\natexlab{a}})\citenamefont
  {Xu}, \citenamefont {Liu}, \citenamefont {Alidoust}, \citenamefont {Neupane},
  \citenamefont {Qian}, \citenamefont {Belopolski}, \citenamefont {Denlinger},
  \citenamefont {Wang}, \citenamefont {Lin}, \citenamefont {Wray} \emph
  {et~al.}}]{xu2012observation}%
  \BibitemOpen
  \bibfield  {author} {\bibinfo {author} {\bibfnamefont {S.-Y.}\ \bibnamefont
  {Xu}}, \bibinfo {author} {\bibfnamefont {C.}~\bibnamefont {Liu}}, \bibinfo
  {author} {\bibfnamefont {N.}~\bibnamefont {Alidoust}}, \bibinfo {author}
  {\bibfnamefont {M.}~\bibnamefont {Neupane}}, \bibinfo {author} {\bibfnamefont
  {D.}~\bibnamefont {Qian}}, \bibinfo {author} {\bibfnamefont {I.}~\bibnamefont
  {Belopolski}}, \bibinfo {author} {\bibfnamefont {J.}~\bibnamefont
  {Denlinger}}, \bibinfo {author} {\bibfnamefont {Y.}~\bibnamefont {Wang}},
  \bibinfo {author} {\bibfnamefont {H.}~\bibnamefont {Lin}}, \bibinfo {author}
  {\bibfnamefont {L.}~\bibnamefont {Wray}},  \emph {et~al.},\ }\href@noop {}
  {\bibfield  {journal} {\bibinfo  {journal} {Nat. Commun.}\ }\textbf {\bibinfo
  {volume} {3}},\ \bibinfo {pages} {1192} (\bibinfo {year}
  {2012}{\natexlab{a}})}\BibitemShut {NoStop}%
\bibitem [{\citenamefont {Liu}\ \emph {et~al.}(2013{\natexlab{a}})\citenamefont
  {Liu}, \citenamefont {Duan},\ and\ \citenamefont {Fu}}]{liu2013two}%
  \BibitemOpen
  \bibfield  {author} {\bibinfo {author} {\bibfnamefont {J.}~\bibnamefont
  {Liu}}, \bibinfo {author} {\bibfnamefont {W.}~\bibnamefont {Duan}}, \ and\
  \bibinfo {author} {\bibfnamefont {L.}~\bibnamefont {Fu}},\ }\href@noop {}
  {\bibfield  {journal} {\bibinfo  {journal} {Phys. Rev. B}\ }\textbf {\bibinfo
  {volume} {88}},\ \bibinfo {pages} {241303} (\bibinfo {year}
  {2013}{\natexlab{a}})}\BibitemShut {NoStop}%
\bibitem [{\citenamefont {Yao}\ and\ \citenamefont
  {Yang}(2015)}]{PhysRevB.92.035132}%
  \BibitemOpen
  \bibfield  {author} {\bibinfo {author} {\bibfnamefont {H.}~\bibnamefont
  {Yao}}\ and\ \bibinfo {author} {\bibfnamefont {F.}~\bibnamefont {Yang}},\
  }\href {\doibase 10.1103/PhysRevB.92.035132} {\bibfield  {journal} {\bibinfo
  {journal} {Phys. Rev. B}\ }\textbf {\bibinfo {volume} {92}},\ \bibinfo
  {pages} {035132} (\bibinfo {year} {2015})}\BibitemShut {NoStop}%
\bibitem [{\citenamefont {Dzyaloshinskii}(1987)}]{dzyaloshinskii1987maximal}%
  \BibitemOpen
  \bibfield  {author} {\bibinfo {author} {\bibfnamefont {I.}~\bibnamefont
  {Dzyaloshinskii}},\ }\href@noop {} {\bibfield  {journal} {\bibinfo  {journal}
  {JETP Lett.}\ }\textbf {\bibinfo {volume} {46}} (\bibinfo {year}
  {1987})}\BibitemShut {NoStop}%
\bibitem [{\citenamefont {Dzyaloshinskii}(1996)}]{dzyaloshinskii1996extended}%
  \BibitemOpen
  \bibfield  {author} {\bibinfo {author} {\bibfnamefont {I.}~\bibnamefont
  {Dzyaloshinskii}},\ }\href@noop {} {\bibfield  {journal} {\bibinfo  {journal}
  {J. Phys. I}\ }\textbf {\bibinfo {volume} {6}},\ \bibinfo {pages} {119}
  (\bibinfo {year} {1996})}\BibitemShut {NoStop}%
\bibitem [{\citenamefont {Baranov}\ \emph {et~al.}(1992)\citenamefont
  {Baranov}, \citenamefont {Chubukov},\ and\ \citenamefont
  {YU.~KAGAN}}]{baranov1992superconductivity}%
  \BibitemOpen
  \bibfield  {author} {\bibinfo {author} {\bibfnamefont {M.}~\bibnamefont
  {Baranov}}, \bibinfo {author} {\bibfnamefont {A.}~\bibnamefont {Chubukov}}, \
  and\ \bibinfo {author} {\bibfnamefont {M.}~\bibnamefont {YU.~KAGAN}},\
  }\href@noop {} {\bibfield  {journal} {\bibinfo  {journal} {IJMPB}\ }\textbf
  {\bibinfo {volume} {6}},\ \bibinfo {pages} {2471} (\bibinfo {year}
  {1992})}\BibitemShut {NoStop}%
\bibitem [{\citenamefont {Bernevig}\ \emph {et~al.}(2006)\citenamefont
  {Bernevig}, \citenamefont {Hughes},\ and\ \citenamefont
  {Zhang}}]{bernevig2006quantum}%
  \BibitemOpen
  \bibfield  {author} {\bibinfo {author} {\bibfnamefont {B.~A.}\ \bibnamefont
  {Bernevig}}, \bibinfo {author} {\bibfnamefont {T.~L.}\ \bibnamefont
  {Hughes}}, \ and\ \bibinfo {author} {\bibfnamefont {S.-C.}\ \bibnamefont
  {Zhang}},\ }\href@noop {} {\bibfield  {journal} {\bibinfo  {journal}
  {Science}\ }\textbf {\bibinfo {volume} {314}},\ \bibinfo {pages} {1757}
  (\bibinfo {year} {2006})}\BibitemShut {NoStop}%
\bibitem [{\citenamefont {Gonz{\'a}lez}\ \emph {et~al.}(1996)\citenamefont
  {Gonz{\'a}lez}, \citenamefont {Guinea},\ and\ \citenamefont
  {Vozmediano}}]{gonzalez1996renormalization}%
  \BibitemOpen
  \bibfield  {author} {\bibinfo {author} {\bibfnamefont {J.}~\bibnamefont
  {Gonz{\'a}lez}}, \bibinfo {author} {\bibfnamefont {F.}~\bibnamefont
  {Guinea}}, \ and\ \bibinfo {author} {\bibfnamefont {M.}~\bibnamefont
  {Vozmediano}},\ }\href@noop {} {\bibfield  {journal} {\bibinfo  {journal}
  {Europhys. Lett.}\ }\textbf {\bibinfo {volume} {34}},\ \bibinfo {pages} {711}
  (\bibinfo {year} {1996})}\BibitemShut {NoStop}%
\bibitem [{\citenamefont {Honerkamp}\ and\ \citenamefont
  {Salmhofer}(2001)}]{honerkamp2001magnetic}%
  \BibitemOpen
  \bibfield  {author} {\bibinfo {author} {\bibfnamefont {C.}~\bibnamefont
  {Honerkamp}}\ and\ \bibinfo {author} {\bibfnamefont {M.}~\bibnamefont
  {Salmhofer}},\ }\href@noop {} {\bibfield  {journal} {\bibinfo  {journal}
  {Phys. Rev. Lett.}\ }\textbf {\bibinfo {volume} {87}},\ \bibinfo {pages}
  {187004} (\bibinfo {year} {2001})}\BibitemShut {NoStop}%
\bibitem [{\citenamefont {McChesney}\ \emph {et~al.}(2010)\citenamefont
  {McChesney}, \citenamefont {Bostwick}, \citenamefont {Ohta}, \citenamefont
  {Seyller}, \citenamefont {Horn}, \citenamefont {Gonz{\'a}lez},\ and\
  \citenamefont {Rotenberg}}]{mcchesney2010extended}%
  \BibitemOpen
  \bibfield  {author} {\bibinfo {author} {\bibfnamefont {J.~L.}\ \bibnamefont
  {McChesney}}, \bibinfo {author} {\bibfnamefont {A.}~\bibnamefont {Bostwick}},
  \bibinfo {author} {\bibfnamefont {T.}~\bibnamefont {Ohta}}, \bibinfo {author}
  {\bibfnamefont {T.}~\bibnamefont {Seyller}}, \bibinfo {author} {\bibfnamefont
  {K.}~\bibnamefont {Horn}}, \bibinfo {author} {\bibfnamefont {J.}~\bibnamefont
  {Gonz{\'a}lez}}, \ and\ \bibinfo {author} {\bibfnamefont {E.}~\bibnamefont
  {Rotenberg}},\ }\href@noop {} {\bibfield  {journal} {\bibinfo  {journal}
  {Phys. Rev. Lett.}\ }\textbf {\bibinfo {volume} {104}},\ \bibinfo {pages}
  {136803} (\bibinfo {year} {2010})}\BibitemShut {NoStop}%
\bibitem [{\citenamefont {Norman}(2011)}]{norman2011challenge}%
  \BibitemOpen
  \bibfield  {author} {\bibinfo {author} {\bibfnamefont {M.~R.}\ \bibnamefont
  {Norman}},\ }\href@noop {} {\bibfield  {journal} {\bibinfo  {journal}
  {Science}\ }\textbf {\bibinfo {volume} {332}},\ \bibinfo {pages} {196}
  (\bibinfo {year} {2011})}\BibitemShut {NoStop}%
\bibitem [{\citenamefont {Mineev}\ \emph {et~al.}(1999)\citenamefont {Mineev},
  \citenamefont {Samokhin},\ and\ \citenamefont
  {Landau}}]{mineev1999introduction}%
  \BibitemOpen
  \bibfield  {author} {\bibinfo {author} {\bibfnamefont {V.~P.}\ \bibnamefont
  {Mineev}}, \bibinfo {author} {\bibfnamefont {K.}~\bibnamefont {Samokhin}}, \
  and\ \bibinfo {author} {\bibfnamefont {L.}~\bibnamefont {Landau}},\
  }\href@noop {} {\emph {\bibinfo {title} {Introduction to unconventional
  superconductivity}}}\ (\bibinfo  {publisher} {CRC Press},\ \bibinfo {year}
  {1999})\BibitemShut {NoStop}%
\bibitem [{\citenamefont {Sigrist}\ and\ \citenamefont
  {Ueda}(1991)}]{sigrist1991phenomenological}%
  \BibitemOpen
  \bibfield  {author} {\bibinfo {author} {\bibfnamefont {M.}~\bibnamefont
  {Sigrist}}\ and\ \bibinfo {author} {\bibfnamefont {K.}~\bibnamefont {Ueda}},\
  }\href@noop {} {\bibfield  {journal} {\bibinfo  {journal} {Rev. Mod. Phys.}\
  }\textbf {\bibinfo {volume} {63}},\ \bibinfo {pages} {239} (\bibinfo {year}
  {1991})}\BibitemShut {NoStop}%
\bibitem [{\citenamefont {Furukawa}\ \emph {et~al.}(1998)\citenamefont
  {Furukawa}, \citenamefont {Rice},\ and\ \citenamefont
  {Salmhofer}}]{furukawa1998truncation}%
  \BibitemOpen
  \bibfield  {author} {\bibinfo {author} {\bibfnamefont {N.}~\bibnamefont
  {Furukawa}}, \bibinfo {author} {\bibfnamefont {T.}~\bibnamefont {Rice}}, \
  and\ \bibinfo {author} {\bibfnamefont {M.}~\bibnamefont {Salmhofer}},\
  }\href@noop {} {\bibfield  {journal} {\bibinfo  {journal} {Phys. Rev. Lett.}\
  }\textbf {\bibinfo {volume} {81}},\ \bibinfo {pages} {3195} (\bibinfo {year}
  {1998})}\BibitemShut {NoStop}%
\bibitem [{\citenamefont {Nandkishore}\ \emph {et~al.}(2012)\citenamefont
  {Nandkishore}, \citenamefont {Levitov},\ and\ \citenamefont
  {Chubukov}}]{nandkishore2012chiral}%
  \BibitemOpen
  \bibfield  {author} {\bibinfo {author} {\bibfnamefont {R.}~\bibnamefont
  {Nandkishore}}, \bibinfo {author} {\bibfnamefont {L.}~\bibnamefont
  {Levitov}}, \ and\ \bibinfo {author} {\bibfnamefont {A.}~\bibnamefont
  {Chubukov}},\ }\href@noop {} {\bibfield  {journal} {\bibinfo  {journal}
  {Nature Phys.}\ }\textbf {\bibinfo {volume} {8}},\ \bibinfo {pages} {158}
  (\bibinfo {year} {2012})}\BibitemShut {NoStop}%
\bibitem [{\citenamefont {Huang}\ \emph
  {et~al.}(2016{\natexlab{a}})\citenamefont {Huang}, \citenamefont {Hsu},
  \citenamefont {Lin}, \citenamefont {Yao},\ and\ \citenamefont
  {Tsai}}]{PhysRevB.93.155108}%
  \BibitemOpen
  \bibfield  {author} {\bibinfo {author} {\bibfnamefont {J.-Q.}\ \bibnamefont
  {Huang}}, \bibinfo {author} {\bibfnamefont {C.-H.}\ \bibnamefont {Hsu}},
  \bibinfo {author} {\bibfnamefont {H.}~\bibnamefont {Lin}}, \bibinfo {author}
  {\bibfnamefont {D.-X.}\ \bibnamefont {Yao}}, \ and\ \bibinfo {author}
  {\bibfnamefont {W.-F.}\ \bibnamefont {Tsai}},\ }\href {\doibase
  10.1103/PhysRevB.93.155108} {\bibfield  {journal} {\bibinfo  {journal} {Phys.
  Rev. B}\ }\textbf {\bibinfo {volume} {93}},\ \bibinfo {pages} {155108}
  (\bibinfo {year} {2016}{\natexlab{a}})}\BibitemShut {NoStop}%
\bibitem [{\citenamefont {Le~Hur}\ and\ \citenamefont
  {Rice}(2009)}]{le2009superconductivity}%
  \BibitemOpen
  \bibfield  {author} {\bibinfo {author} {\bibfnamefont {K.}~\bibnamefont
  {Le~Hur}}\ and\ \bibinfo {author} {\bibfnamefont {T.~M.}\ \bibnamefont
  {Rice}},\ }\href@noop {} {\bibfield  {journal} {\bibinfo  {journal} {Ann.
  Phys}\ }\textbf {\bibinfo {volume} {324}},\ \bibinfo {pages} {1452} (\bibinfo
  {year} {2009})}\BibitemShut {NoStop}%
\bibitem [{\citenamefont {Kundu}\ and\ \citenamefont
  {Tripathi}(2017)}]{kundu2017role}%
  \BibitemOpen
  \bibfield  {author} {\bibinfo {author} {\bibfnamefont {S.}~\bibnamefont
  {Kundu}}\ and\ \bibinfo {author} {\bibfnamefont {V.}~\bibnamefont
  {Tripathi}},\ }\href@noop {} {\bibfield  {journal} {\bibinfo  {journal}
  {arXiv preprint arXiv:1704.07437}\ } (\bibinfo {year} {2017})}\BibitemShut
  {NoStop}%
\bibitem [{\citenamefont {Serbyn}\ and\ \citenamefont {Fu}(2014)}]{article}%
  \BibitemOpen
  \bibfield  {author} {\bibinfo {author} {\bibfnamefont {M.}~\bibnamefont
  {Serbyn}}\ and\ \bibinfo {author} {\bibfnamefont {L.}~\bibnamefont {Fu}},\
  }\bibfield  {booktitle} {\emph {\bibinfo {booktitle} {Physical Review B}},\
  }\href@noop {} {\ \textbf {\bibinfo {volume} {90}} (\bibinfo {year}
  {2014})}\BibitemShut {NoStop}%
\bibitem [{\citenamefont {Huang}\ \emph
  {et~al.}(2016{\natexlab{b}})\citenamefont {Huang}, \citenamefont {Lin},
  \citenamefont {Wang}, \citenamefont {Bansil},\ and\ \citenamefont
  {Tsai}}]{huang2016hedgehog}%
  \BibitemOpen
  \bibfield  {author} {\bibinfo {author} {\bibfnamefont {C.-Y.}\ \bibnamefont
  {Huang}}, \bibinfo {author} {\bibfnamefont {H.}~\bibnamefont {Lin}}, \bibinfo
  {author} {\bibfnamefont {Y.~J.}\ \bibnamefont {Wang}}, \bibinfo {author}
  {\bibfnamefont {A.}~\bibnamefont {Bansil}}, \ and\ \bibinfo {author}
  {\bibfnamefont {W.-F.}\ \bibnamefont {Tsai}},\ }\href@noop {} {\bibfield
  {journal} {\bibinfo  {journal} {Physical Review B}\ }\textbf {\bibinfo
  {volume} {93}},\ \bibinfo {pages} {205304} (\bibinfo {year}
  {2016}{\natexlab{b}})}\BibitemShut {NoStop}%
\bibitem [{\citenamefont {Liu}\ \emph {et~al.}(2013{\natexlab{b}})\citenamefont
  {Liu}, \citenamefont {Hsieh}, \citenamefont {Wei}, \citenamefont {Duan},
  \citenamefont {Moodera},\ and\ \citenamefont {Fu}}]{liu2013spin}%
  \BibitemOpen
  \bibfield  {author} {\bibinfo {author} {\bibfnamefont {J.}~\bibnamefont
  {Liu}}, \bibinfo {author} {\bibfnamefont {T.~H.}\ \bibnamefont {Hsieh}},
  \bibinfo {author} {\bibfnamefont {P.}~\bibnamefont {Wei}}, \bibinfo {author}
  {\bibfnamefont {W.}~\bibnamefont {Duan}}, \bibinfo {author} {\bibfnamefont
  {J.}~\bibnamefont {Moodera}}, \ and\ \bibinfo {author} {\bibfnamefont
  {L.}~\bibnamefont {Fu}},\ }\href@noop {} {\bibfield  {journal} {\bibinfo
  {journal} {arXiv preprint arXiv:1310.1044}\ } (\bibinfo {year}
  {2013}{\natexlab{b}})}\BibitemShut {NoStop}%
\bibitem [{\citenamefont {Fang}\ \emph {et~al.}(2014)\citenamefont {Fang},
  \citenamefont {Gilbert},\ and\ \citenamefont
  {Bernevig}}]{PhysRevLett.112.046801}%
  \BibitemOpen
  \bibfield  {author} {\bibinfo {author} {\bibfnamefont {C.}~\bibnamefont
  {Fang}}, \bibinfo {author} {\bibfnamefont {M.~J.}\ \bibnamefont {Gilbert}}, \
  and\ \bibinfo {author} {\bibfnamefont {B.~A.}\ \bibnamefont {Bernevig}},\
  }\href {\doibase 10.1103/PhysRevLett.112.046801} {\bibfield  {journal}
  {\bibinfo  {journal} {Phys. Rev. Lett.}\ }\textbf {\bibinfo {volume} {112}},\
  \bibinfo {pages} {046801} (\bibinfo {year} {2014})}\BibitemShut {NoStop}%
\bibitem [{\citenamefont {Xu}\ \emph {et~al.}(2012{\natexlab{b}})\citenamefont
  {Xu}, \citenamefont {Neupane}, \citenamefont {Liu}, \citenamefont {Zhang},
  \citenamefont {Richardella}, \citenamefont {Wray}, \citenamefont {Alidoust},
  \citenamefont {Leandersson}, \citenamefont {Balasubramanian}, \citenamefont
  {S{\'a}nchez-Barriga} \emph {et~al.}}]{xu2012hedgehog}%
  \BibitemOpen
  \bibfield  {author} {\bibinfo {author} {\bibfnamefont {S.-Y.}\ \bibnamefont
  {Xu}}, \bibinfo {author} {\bibfnamefont {M.}~\bibnamefont {Neupane}},
  \bibinfo {author} {\bibfnamefont {C.}~\bibnamefont {Liu}}, \bibinfo {author}
  {\bibfnamefont {D.}~\bibnamefont {Zhang}}, \bibinfo {author} {\bibfnamefont
  {A.}~\bibnamefont {Richardella}}, \bibinfo {author} {\bibfnamefont {L.~A.}\
  \bibnamefont {Wray}}, \bibinfo {author} {\bibfnamefont {N.}~\bibnamefont
  {Alidoust}}, \bibinfo {author} {\bibfnamefont {M.}~\bibnamefont
  {Leandersson}}, \bibinfo {author} {\bibfnamefont {T.}~\bibnamefont
  {Balasubramanian}}, \bibinfo {author} {\bibfnamefont {J.}~\bibnamefont
  {S{\'a}nchez-Barriga}},  \emph {et~al.},\ }\href@noop {} {\bibfield
  {journal} {\bibinfo  {journal} {arXiv preprint arXiv:1212.3382}\ } (\bibinfo
  {year} {2012}{\natexlab{b}})}\BibitemShut {NoStop}%
\bibitem [{\citenamefont {Chubukov}(2009)}]{chubukov2009renormalization}%
  \BibitemOpen
  \bibfield  {author} {\bibinfo {author} {\bibfnamefont {A.}~\bibnamefont
  {Chubukov}},\ }\href@noop {} {\bibfield  {journal} {\bibinfo  {journal}
  {Physica C}\ }\textbf {\bibinfo {volume} {469}},\ \bibinfo {pages} {640}
  (\bibinfo {year} {2009})}\BibitemShut {NoStop}%
\bibitem [{\citenamefont {Michaeli}\ and\ \citenamefont
  {Fu}(2012)}]{PhysRevLett.109.187003}%
  \BibitemOpen
  \bibfield  {author} {\bibinfo {author} {\bibfnamefont {K.}~\bibnamefont
  {Michaeli}}\ and\ \bibinfo {author} {\bibfnamefont {L.}~\bibnamefont {Fu}},\
  }\href {\doibase 10.1103/PhysRevLett.109.187003} {\bibfield  {journal}
  {\bibinfo  {journal} {Phys. Rev. Lett.}\ }\textbf {\bibinfo {volume} {109}},\
  \bibinfo {pages} {187003} (\bibinfo {year} {2012})}\BibitemShut {NoStop}%
\bibitem [{\citenamefont {Nagai}(2015)}]{nagai2015robust}%
  \BibitemOpen
  \bibfield  {author} {\bibinfo {author} {\bibfnamefont {Y.}~\bibnamefont
  {Nagai}},\ }\href@noop {} {\bibfield  {journal} {\bibinfo  {journal}
  {Physical Review B}\ }\textbf {\bibinfo {volume} {91}},\ \bibinfo {pages}
  {060502} (\bibinfo {year} {2015})}\BibitemShut {NoStop}%
\bibitem [{\citenamefont {Mackenzie}\ \emph {et~al.}(1998)\citenamefont
  {Mackenzie}, \citenamefont {Haselwimmer}, \citenamefont {Tyler},
  \citenamefont {Lonzarich}, \citenamefont {Mori}, \citenamefont {Nishizaki},\
  and\ \citenamefont {Maeno}}]{PhysRevLett.80.161}%
  \BibitemOpen
  \bibfield  {author} {\bibinfo {author} {\bibfnamefont {A.~P.}\ \bibnamefont
  {Mackenzie}}, \bibinfo {author} {\bibfnamefont {R.~K.~W.}\ \bibnamefont
  {Haselwimmer}}, \bibinfo {author} {\bibfnamefont {A.~W.}\ \bibnamefont
  {Tyler}}, \bibinfo {author} {\bibfnamefont {G.~G.}\ \bibnamefont
  {Lonzarich}}, \bibinfo {author} {\bibfnamefont {Y.}~\bibnamefont {Mori}},
  \bibinfo {author} {\bibfnamefont {S.}~\bibnamefont {Nishizaki}}, \ and\
  \bibinfo {author} {\bibfnamefont {Y.}~\bibnamefont {Maeno}},\ }\href
  {\doibase 10.1103/PhysRevLett.80.161} {\bibfield  {journal} {\bibinfo
  {journal} {Phys. Rev. Lett.}\ }\textbf {\bibinfo {volume} {80}},\ \bibinfo
  {pages} {161} (\bibinfo {year} {1998})}\BibitemShut {NoStop}%
\bibitem [{\citenamefont {Das}\ \emph {et~al.}(2016)\citenamefont {Das},
  \citenamefont {Aggarwal}, \citenamefont {Roychowdhury}, \citenamefont
  {Aslam}, \citenamefont {Gayen}, \citenamefont {Biswas},\ and\ \citenamefont
  {Sheet}}]{das2016unexpected}%
  \BibitemOpen
  \bibfield  {author} {\bibinfo {author} {\bibfnamefont {S.}~\bibnamefont
  {Das}}, \bibinfo {author} {\bibfnamefont {L.}~\bibnamefont {Aggarwal}},
  \bibinfo {author} {\bibfnamefont {S.}~\bibnamefont {Roychowdhury}}, \bibinfo
  {author} {\bibfnamefont {M.}~\bibnamefont {Aslam}}, \bibinfo {author}
  {\bibfnamefont {S.}~\bibnamefont {Gayen}}, \bibinfo {author} {\bibfnamefont
  {K.}~\bibnamefont {Biswas}}, \ and\ \bibinfo {author} {\bibfnamefont
  {G.}~\bibnamefont {Sheet}},\ }\href@noop {} {\bibfield  {journal} {\bibinfo
  {journal} {Appl. Phys. Lett.}\ }\textbf {\bibinfo {volume} {109}},\ \bibinfo
  {pages} {132601} (\bibinfo {year} {2016})}\BibitemShut {NoStop}%
\bibitem [{\citenamefont {He}\ \emph {et~al.}(2017)\citenamefont {He},
  \citenamefont {Pan}, \citenamefont {Stern}, \citenamefont {Burks},
  \citenamefont {Che}, \citenamefont {Yin}, \citenamefont {Wang}, \citenamefont
  {Lian}, \citenamefont {Zhou}, \citenamefont {Choi} \emph
  {et~al.}}]{he2017chiral}%
  \BibitemOpen
  \bibfield  {author} {\bibinfo {author} {\bibfnamefont {Q.~L.}\ \bibnamefont
  {He}}, \bibinfo {author} {\bibfnamefont {L.}~\bibnamefont {Pan}}, \bibinfo
  {author} {\bibfnamefont {A.~L.}\ \bibnamefont {Stern}}, \bibinfo {author}
  {\bibfnamefont {E.~C.}\ \bibnamefont {Burks}}, \bibinfo {author}
  {\bibfnamefont {X.}~\bibnamefont {Che}}, \bibinfo {author} {\bibfnamefont
  {G.}~\bibnamefont {Yin}}, \bibinfo {author} {\bibfnamefont {J.}~\bibnamefont
  {Wang}}, \bibinfo {author} {\bibfnamefont {B.}~\bibnamefont {Lian}}, \bibinfo
  {author} {\bibfnamefont {Q.}~\bibnamefont {Zhou}}, \bibinfo {author}
  {\bibfnamefont {E.~S.}\ \bibnamefont {Choi}},  \emph {et~al.},\ }\href@noop
  {} {\bibfield  {journal} {\bibinfo  {journal} {Science}\ }\textbf {\bibinfo
  {volume} {357}},\ \bibinfo {pages} {294} (\bibinfo {year}
  {2017})}\BibitemShut {NoStop}%
\end{thebibliography}

\end{document}